\documentclass[]{interact}

\usepackage{epstopdf}
\usepackage[caption=false]{subfig}

\usepackage[natbibapa,nodoi]{apacite}
\usepackage{amsmath,amsfonts,amsthm,amssymb}
\usepackage{algorithm}
\usepackage{algpseudocode}
\usepackage{array}
\usepackage{textcomp}
\usepackage{stfloats}
\usepackage{url}
\usepackage{verbatim}
\usepackage{multirow}
\usepackage{multicol}
\usepackage{tabularx}[color]
\usepackage[dvipsnames]{xcolor}
\usepackage{graphicx}
\usepackage{color, colortbl}
\usepackage{sidecap}
\usepackage{xcolor}
\usepackage{todonotes}

\def\bs{\boldsymbol}

\theoremstyle{plain}
\newtheorem{theorem}{Theorem}[section]

\theoremstyle{definition}
\newtheorem{definition}[theorem]{Definition}

\theoremstyle{remark}

\begin{document}


\title{Efficient and Robust Remote Sensing Image Denoising Using Randomized Approximation of Geodesics' Gramian on the Manifold Underlying the Patch Space}

\author{
\name{Kelum~Gajamannage\textsuperscript{a}\thanks{CONTACT K.~Gajamannage. Email: kelum.gajamannage@uri.edu}, Dilhani I. Jayathilake\textsuperscript{b}, and Maria Vasilyeva\textsuperscript{c}}
\affil{\textsuperscript{a}Department of Mathematics and Applied Mathematical Sciences, University of Rhode Island, Kingston, RI 02881, USA; 
\textsuperscript{b}School of Computing and Engineering, Quinnipiac University, Hamden, CT 06518, USA;
\textsuperscript{c}Department of Mathematics and Statistics, Texas A\&M University - Corpus Christi, Corpus Christi, TX 78412, USA}
}
\maketitle

\begin{abstract}
Remote sensing images are widely utilized in many disciplines such as feature recognition and scene semantic segmentation. However, due to environmental factors and the issues of the imaging system, the image quality is often degraded which may impair subsequent visual tasks. Even though denoising remote sensing images plays an essential role before applications, the current denoising algorithms fail to attain optimum performance since these images possess complex features in the texture. Denoising frameworks based on artificial neural networks have shown better performance; however, they require exhaustive training with heterogeneous samples that extensively consume resources like power, memory, computation, and latency. Thus, here we present a computationally efficient and robust remote sensing image denoising method that doesn’t require additional training samples. This method partitions patches of a remote-sensing image in which a low-rank manifold, representing the noise-free version of the image, underlies the patch space. An efficient and robust approach to revealing this manifold is a randomized approximation of the singular value spectrum of the geodesics' Gramian matrix of the patch space. The method asserts a unique emphasis on each color channel during denoising so the three denoised channels are merged to produce the final image.
\end{abstract}


\section{Introduction}\label{sec:int}
Remote sensing is the process of detecting and monitoring the physical characteristics of an area by measuring its reflected and emitted radiation at a distance \citep{campbell2011introduction}. Remote sensing allows us to see much more than we can see from proximity and sometimes it is the only option for imaging unreached extends such as celestial objects and ocean flow. Some specific uses of remote sensing include mapping extensive forest fires for rangers to efficiently act for extinguishing \citep{kusangaya2015remote}, sensing clouds to help predict the weather \citep{matsunobu2021cloud}, and tracking the land cover of forests versus farmlands/cities to predict global warming \citep{lossou2019monitoring}. Due to its immense applicability, the development of remote technology that generates highly informative and precise images has been emerging for decades \citep{liu2019remote}. Although high spatial and spectral resolution is desired in remote sensing images they might be distorted by noise due to factors like device failures and environmental conditions \citep{GajamannageGGDwtNoise}. A vital preprocessing step is required for denoising those images to enhance the quality. Both algebraic and Artificial Neural Network (ANN) methods have been developed, and have shown good performance, for remote sensing image denoising. However, while ANN-based frameworks require a significant amount of training samples which makes the process less efficient, most of the algebraic methods require unique tuning of hyperparameters for each subject of interest which makes the process less robust \citep{gajamannage2022reconstruction}. With all in mind, here we present a computationally efficient and robust remote sensing image denoising method that is capable of reconstructing accurate images without any additional training images.

Among algebraic denoising frameworks, sparse 3-D transform-domain collaborative filtering \citep{Dabov2007}, abbreviated as BM3D, sparse and redundant representations over learned dictionaries \citep{Elad2006}, abbreviated as KSVD, are known to be the most effective. BM3D consists of a two-stage non-local collaborative filtering process in the transform domain which stacks similar patches into 3D groups by block matching and then transforms these 3D groups into the wavelet domain \citep{Dabov2007}. Hard thresholding or Wiener filtering with coefficients is employed in the wavelet domain followed by applying an inverse transformation of coefficients. Finally, the denoised version of the image is constructed by aggregating all the estimated patches. However, when the noise increases gradually, the denoising performance of BM3D decreases greatly and artifacts are introduced, especially around uniform areas \citep{Fan2019}. KSVD is based on sparse and redundant representations over trained dictionaries in which the training is performed either using the corrupted image itself or training on a corpus of high-quality image databases. For an input threshold of the sparsity, KSVD performs denoising using a global image prior estimated by a Bayesian reconstruction framework that forces sparsity over patches in every location in the image. However, this method has computational deficiencies since both the choice of an appropriate dictionary for a dataset is a non-convex problem and the implementation of KSVD is an iterative approach that does not guarantee finding the global optimum \citep{Rubinstein2010}. 

Attention-guided CNN for image denoising \citep{tian2020attention}, abbreviated as ADNet, and feed-forward denoising convolutional neural networks \citep{zhang2017beyond}, abbreviated as DnCNN, are known to be among the most famous ANN-based denoising frameworks. ADNet includes four blocks for image denoising: 1) a sparse block making a tradeoff between performance and efficiency by using dilated and common convolutions, 2) a feature enhancement block integrating global and local feature information via a long path to enhance the expressive ability, 3) an attention block extracting the noise information hidden in the complex background, and 4) a reconstruction block constructing the clean image through the obtained noise mapping and the given noisy image. DnCNN utilizes residual learning and batch normalization to both speed up the training process and boost the denoising performance. DnCNN is capable of handling Gaussian denoising with unknown noise levels and is implemented by benefiting from GPU computing. With the residual learning strategy, DnCNN implicitly removes the latent clean image in the hidden layers. This property motivates us to train a single DnCNN model to tackle several general image denoising tasks, such as Gaussian denoising, single image super-resolution, and JPEG image deblocking. However, ANN-based denoising frameworks may be inefficient in providing the original structural features at increasing noise levels and also require a significant amount of training samples for sustainable performance \citep{kirik2024quantitative}.

Our remote sensing image denoising method presented here is both efficient and robust as it uses approximated singular vectors (SVs) of the Gramian matrix of geodesic distances. For a given Red-Green-Blue (RGB) remote sensing image, our method applies a two-dimensional image denoising framework for each color channel separately and then combines denoised channels to reconstruct the image. For that, we partition each color channel of the noisy image into partially overlapping moving square patches with a known length, say $\rho$, such that each patch is centered at one unique pixel of the image. For a given color channel, each patch is a point in a $\rho^2$-dimensional space where a low-dimensional manifold underlies \citep{gajamannage2015a, gajamannage2015identifying}. These manifold representations are similar to the wavelet domain representation of BM3D and BWD, and the redundant dictionary representation of KSVD, where the image features are concentrated. Revealing this hidden manifold helps better explain the geometry of the patch-set and then helps identify the features in the image. Learning such manifolds is the salient step in the discipline of \emph{Dimensionality Reduction} from where we borrow the basic concept for the proposed method for denoising images. In the context of image denoising, the process of projecting the high-dimensional data of the patch space into a low-dimensional manifold is technically the same as eliminating the noise in the image since extra dimensions mostly represent noise and minor information. This projection is conducted using the Gramian matrix of geodesic distances \citep{SGE,gajamannage2021}. Due to this manifold approach, our method can be considered to be a non-local denoising method that performs denoising in the patch space rather than in the image domain to preserve the features across the entire image.

For a given RGB image of size $n\times n\times 3$, we compute the geodesic distance between each pair of patches on each of the three manifolds using a shortest paths algorithm and formulate a geodesic distance matrix of size $n^2\times n^2$ for each of the manifolds representing each color channel. Each of the geodesic matrices is transformed into its Gramian matrix of the same size and the SVs corresponding to prominent singular values of each of the $n^2\times n^2$ Gramian matrix is used to denoise each channel. Since this framework is an efficient three-channel image denoising method based on geodesics' Gramian matrix, we name it Efficient Geodesic Gramian Denoising, abbreviated as EGGD. The computational complexity of the Singular Value Decomposition (SVD) to compute SVs is as high as $\mathcal{O}(n^6)$ for each channel \citep{gajamannage2022efficient}; thus, we rely on the singular vector approximation scheme named Randomized Singular Value Decomposition (RSVD) which approximates SVs efficiently. With RSVD, the computational complexity of EGGD is reduced to $\mathcal{O}\left(n^4 + k^2n^2\right)$ where $k$ is the required number of prominent SVs. Since $k$ is significantly less than $n$, the ultimate computational complexity of EGGD becomes $\mathcal{O}(n^4)$.

EGGD transforms an RGB image into a three-channel, Luminance, Chrominance-blue, and Chrominance-red (YCbCr), color spectrum. As we will see in the sequel, each of the channels, R, G, and B, possess rich information; thus, denoising the channels R, G, and B requires equally high computation emphasis on each of the channels. However, channel Y of the YCbCr representation possesses the greatest information while the other two channels Cb and Cr contain only minor information. Thus, we perform denoising on the YCbCr spectrum and assert more emphasis only on the Y channel during denoising to attain greater efficiency. Since remote sensing images contain more information than regular images, we also give adequate emphasis to both Cb and Cr channels for the superior quality of image denoising. 

The rest of this article is structured as follows: Sec.~\ref{sec:cont} states the novelty and contribution of this work; Sec.~\ref{sec:met} presents methods that include RSVD, denoising of a single-color channel, and three-color channel denoising; Sec.~\ref{sec:per} presents performance analysis based on a detailed comparison of EGGD with two algebraic denoising frameworks, BM3D and KSVD, and two ANN denoising frameworks, ADNet and DnCNN; and the discussion along with conclusions is presented in Sec.~\ref{sec:con}.

\subsection{Contributions}\label{sec:cont}
Our proposed denoising method, EGGD, makes the following contributions to the literature:
\begin{itemize}
\item EGGD is an algebraic framework that doesn't require any pre-training like ANN-based frameworks; thus, EGGD requires limited resources like power, memory, computation, and latency.
\item EGGD performs denoising in the YCbCr color spectrum rather than in the RGB spectrum to attain greater denoising efficiency and accuracy.
\item EGGD utilizes a few SVs of the Gramian matrix of geodesic distances between patches for denoising, which helps both retain essential image features and remove image noise, with less computational cost.
\item EGGD approximates SVs of the Gramian matrix using Randomized SVD which reduces the computational complexity by $n^2$-times where $n$ is the length of the squared image.
\item EGGD is a non-local patch-based denoising scheme in which non-local property helps preserve the features across the entire image rather than working locally, and patch-based property helps preserve image smoothness, fine image details, and sharp edges than those of pixel-based methods.
\item In contrast to the state-of-the-art non-local denoising methods that are well known to have many parameters, EGGD possesses only three parameters that can easily be pre-determined.
\end{itemize}

\section{Methods}\label{sec:met}
The computational complexity of the SVD to compute SVs is $\mathcal{O}(n^6)$ for each channel of an RGB remote sensing image of size $n\times n\times 3$; thus, we approximate SVs by RSVD which only possesses the complexity of $\mathcal{O}(n^4 + k^2n^2)$, where $k$ is the required number of prominent SVs. This ultimately becomes $\mathcal{O}(n^4)$ since $k^2n^2$ is less influential with a significantly smaller $k$ than $n$. The EGGD framework is implemented by combining three channels each denoised using approximated SVs of the geodesics' Gramian matrix of its own channel. Thus, our image denoising framework has three main parts: approximation of SVs of a matrix using RSVD, single-channel image denoising using approximated SVs of the geodesics' Gramian matrix, and three-channel extension of the single-channel denoising framework.

\subsection{Approximation of SVs of a matrix using randomized singular value decomposition}\label{sec:apr}
Randomized singular value decomposition \citep{halko2011finding}, combines probability theory with numerical linear algebra to develop an efficient, unbiased, and randomized algorithm to approximate SVD \citep{gajamannage2022efficient}. RSVD assumes the data matrix is low-rank and efficiently approximates $L$ number of orthonormal vectors that span the range of the given data matrix $\mathcal{G}\in \mathbb{R}^{m\times n}$ so that such vectors enable an efficient approximation for SVD  of $\mathcal{G}$. RSVD undergoes a two-stage computational process: 1) approximation of orthonormal basis, denoted as $Q$, for the range of $\mathcal{G}$; and 2) approximation of SVD of $\mathcal{G}$ using such orthonormal basis $Q$. Randomness only occurs in Step 1 and Step 2 is deterministic for a given $Q$. 

\begin{itemize}
\item \textbf{First stage}
Here, the goal is to generate an orthonormal matrix $Q_{n\times L}$ that consists of as few columns as possible such that
\begin{eqnarray}\label{eq:projector}
\mathcal{G} \approx QQ^T \mathcal{G},
\end{eqnarray}
where $Q^T$ denotes the transpose of $Q$. First, draw $L$ Gaussian random columns $\bs{\omega}_1$, $\bs{\omega}_2$, \dots, $\bs{\omega}_L$ from $\mathcal{G}_{m\times n}$ and project them using the linear map $\mathcal{G}$ such that $Y_{m\times L}=\mathcal{G}\Omega$ where $\Omega_{n\times L}=[\bs{\omega}_1, \bs{\omega}_2, \dots, \bs{\omega}_L]$. Then, compute the orthonormal matrix $Q$ by using $QR$ factorization of $Y$ such that $Y=QR$. Columns of the matrix $Q$ represent an orthonormal basis for the range of $Y$ and then for $\mathcal{G}$. While having a few columns in the basis matrix $Q$ increases the efficiency of the approximation, having more columns in it increases the accuracy.
\\
\item \textbf{Second stage}
Since the number of columns $L$ of the matrix $Q$ is significantly less than both the dimensions of $\mathcal{G}$, it is efficient to compute SVD on $Q$. Let,
\begin{eqnarray}\label{eq:B}
B_{L\times n} = Q^T \mathcal{G},
\end{eqnarray}
where $B$ consists of only $L$-many rows; then, the SVD of $B$ is efficiently computed using Def.~\ref{def:svd} as
\begin{eqnarray}\label{eq:W}
B = W \Sigma V^T,
\end{eqnarray}
where the columns of both $W_{L\times L}$ and $V_{n\times L}$ are orthonormal, and $\Sigma_{L \times L}$ is a diagonal matrix whose entries are all non-negative. Let,  
\begin{eqnarray}\label{eq:U}
U_{m\times L} = QW
\end{eqnarray}
and combining Eqns.~\eqref{eq:projector}--\eqref{eq:U} yields SVD of $\mathcal{G}$ such that
\begin{eqnarray}\label{eq:comb}
\mathcal{G} \approx QQ^T \mathcal{G} = QB = Q W \Sigma V^T =  U \Sigma V^T.
\end{eqnarray}
\end{itemize}

\begin{definition}\label{def:svd}
Let, $Diag\left(\sigma_1, \dots, \sigma_{\min(m,n)}\right)$ denotes a diagonal matrix constructed with the vector $\left(\sigma_1, \dots, \sigma_{\min(m,n)}\right)$ as its diagonal; $A\in\mathbb{R}^{m\times n}$ be any matrix; and $U_{m\times m}$ and $V_{n\times n}$ are unitary matrices such that $U^TU=I$ and $V^TV=I$, respectively. Then, the SVD of $A$ is $A=U\Sigma V^T$, where $\Sigma_{m\times n}= $$Diag\left(\sigma_1, \dots \sigma_{\min(m,n)}\right)$ with $\sigma_1\ge\dots\ge\sigma_{\min(m,n)}\ge0$. Here, for $l=1 ,\dots, n$, the column vector $\mathbf{v}_l$ represents the $l$-th right singular vector of $A$ such that $V=[\mathbf{v}_1,\dots,\mathbf{v}_l,\dots,\mathbf{v}_{n}]$, \citep{Golub1970}. 
\end{definition}

The RSVD scheme is presented in Alg.~\ref{alg:rsvd}. Since $L$ is significantly less than both the dimensions of $\mathcal{G}$, the matrices in the RSVD scheme have substantially fewer entries than that of $\mathcal{G}$. This RSVD scheme is used to approximate the SVs of the geodesics' Gramian matrix of the patch-space in Sec.~\ref{sec:1ch}. 

\begin{algorithm}[H]
\caption{Randomized Singular Value Decomposition (RSVD) \citep{halko2011finding}.\\
Inputs: data matrix ($\mathcal{G}\in \mathbb{R}^{m\times m}$) and the number of desired singular triplets ($L$).\\
Outputs: approximated singular triplets $\{\sigma_l, \mathbf{u}_l,\mathbf{v}_l\}, l=1,...,L$.}
\label{alg:rsvd}
\begin{algorithmic}[1]
\item[] {\bf \hspace{-0.6cm}First stage:}
\State Create an ($m\times L$)-dimensional projection matrix $\Omega$ by drawing $L$ columns of $\mathcal{G}$ randomly.
\State Compute the projection $Y_{m\times L} = \mathcal{G} \Omega$.
\State Generate an orthonormal matrix $Q_{m\times L}$ using $QR$ factorization, i.e., $Y=Q_{m\times L}R_{L\times L}$. 
\vspace{5mm}
\item[]{\bf \hspace{-0.6cm}Second stage:}
\State Compute $B_{L\times n} = Q^T \mathcal{G}$
\State Perform SVD of the matrix $B = W \Sigma V^T$, where $W\in\mathbb{R}^{L\times L}$, $\Sigma \in\mathbb{R}^{L\times L}$, and $V\in\mathbb{R}^{n\times L}$
\State Set $U_{m\times L} = QW$
\State Return $\{\sigma_l, \mathbf{u}_l, \mathbf{v}_l\}, l=1,...,L$, where $\sigma_l$, $\mathbf{u}_l$, and $\mathbf{v}_l$ are the $l$-th diagonal entry of $\Sigma$, $l$-th column of $U$, and $l$-th column of $V^T$, respectively. 
\end{algorithmic}
\end{algorithm}

\subsection{Single-channel image denoising using SVs of the geodesics' Gramian matrix}\label{sec:1ch}
The single-channel remote sensing image denoising scheme consists of five steps. First, the input noisy image, denoted as $\mathcal{U}_{n\times n}$, is partitioned into overlapping square patches, denoted as $\bs{u}(\bs{x}_{ij})$'s; $i,j=1,\dots,n$, each with a fixed odd length, denoted as $\rho$, such that each patch is centered at each pixel of the image. For $k=n(i-1)+j$ and $1\le k \le n^2$, we write $\bs{u}(\bs{x}_{k})$ instead of $\bs{u}(\bs{x}_{ij})$ on some occasions for simplicity. Each patch $\bs{u}(\bs{x}_{k})$; $i=1,\dots,n^2$ can be represented as a $\rho^2$-dimensional point. A low-dimensional manifold underlies in this high-dimensional data cloud representing the set of patches of this image. Second, we create a graph structure, denoted as $G(V,E)$, using the neighbor search algorithm in \cite{agarwal1999geometric}. Here, 1) the points $\{\bs{u}(\bs{x}_{k})\vert k=1,\dots,n^2\}$ are treated as vertices and denoted by $V$; and, 2) the edge set $E$ is defined by the following process: for a given neighborhood parameter $\delta$, we join each vertex $\bs{u}(\bs{x}_{k})$ to its $\delta$ nearest neighbors such that the weight of the edge between the vertex $\bs{u}(\bs{x}_{k})$ and its given neighbor vertex, denoted as $\bs{u}(\bs{x}_{k'})$, is the Euclidean distance between them, denoted as $d(k, k')$, where
\begin{equation}\label{eqn:eudist}
d(k, k')=\|\bs{u}(\bs{x}_{k})-\bs{u}(\bs{x}_{k'})\|_2.
\end{equation}
Then, the geodesic distance between two nodes on this network representing two patches is approximated by the shortest path distance between those nodes according to Floyd’s algorithm \citep{floyd1962algorithm}.

In the third step, the matrix of geodesic distances, denoted as $\mathcal{D} \in\mathbb{R}^{n^2\times n^2}_{\ge0}$, is transformed into a Gramian matrix, denoted as $\mathcal{G}_{n^2 \times n^2}$, using
\begin{equation}\label{eqn:gram}
\mathcal{G}[i,j]=-\frac{1}{2}\big[\mathcal{D}[i,j]-\mu_i(\mathcal{D}) -\mu_j(\mathcal{D})+\mu(\mathcal{D})\big],
\end{equation}
where $\mu_i(\mathcal{D})$ is the mean of the $i$-th row of the matrix $\mathcal{D}$, $\mu_j(\mathcal{D})$ is the mean of the $j$-th column of that matrix, and $\mu(\mathcal{D})$ is the mean of the full matrix \citep{lee2004nonlinear}. Fourth, EGGD denoises the noisy patches $\bs{u}_{k}$ using only a few, denoted as $L$, prominent right SVs, defined as $\mathbf{v}_l$'s where $l=1, \dots, L$, of the Gramian matrix since SVs of the prominent singular values only represent image features excluding the image noise. For that, we input $\mathcal{G}$ into the RSVD algorithm in Alg.~\ref{alg:rsvd} and approximate $L$ SVs corresponding to the prominent singular values. The noise-free representation of the noisy patch $\bs{u}_{k}$ is denoted as $\tilde{\bs{u}}_{k}$ which EGGD produces by
\begin{equation}\label{eqn:denoise}
\tilde{\bs{u}}(\bs{x}_{k}) = \sum^{L}_{l=1} \langle \bs{u}(\bs{x}_{k}), \mathbf{v}_l \rangle \mathbf{v}_l,
\end{equation}
where $\langle \cdot, \cdot \rangle$ represents the inner product operator and $L$ is the required number of prominent SVs

In the fifth step, EGGD constructs the noise-free image from the denoised patches. For that, it estimates each $k$-th pixel's intensity of the image by using the pixels of $\rho^2$ overlapping patches coincident with this $k$-th pixel. Thus, for the sake of convenient interpretation, we introduce a new index $t_n$ for each pixel $\bs{x}_t \in \mathcal{N}(\bs{x}_k)$, where
\begin{equation}
\mathcal{N}(\bs{x}_{k})=\{\bs{x}_{t} \ \vert \ \ \|\bs{x}_{k}-\bs{x}_{t}\|_{\infty}\le \rho/2\},
\end{equation}
such that the physical location of the pixel of the patch $\tilde{\bs{u}}(\bs{x}_t)$ at the new index, denoted as $[\tilde{\bs{u}}(\bs{x}_t)]_{t_n}$, is the same physical location of the pixel $\bs{x}_k$. Then, all these estimations are combined using a moving least square approximation according to Shepard’s method \citep{Shepard1968}, as
\begin{equation}\label{eqn:merge}
\tilde{\mathcal{U}}(\bs{x}_{k})=\sum_{\bs{x}_t \in \mathcal{N}(\bs{x}_{k})} \Gamma (\bs{x}_{k},\bs{x}_t) [\tilde{\bs{u}}(\bs{x}_t)]_{t_n},
\end{equation}
\citep{Meyer2014}, where the weights are given by
\begin{equation}\label{eqn:shepard_weight}
\Gamma(\bs{x}_{k},\bs{x}_t)=\frac{e^{-\|\bs{x}_{k}-\bs{x}_t\|^2}}{\sum_{\bs{x}_{t'}\in \mathcal{N}(\bs{x}_{k})} e^{-\|\bs{x}_{k}-\bs{x}_{t'}\|^2}}.
\end{equation}
This weighting factor weights nearby pixels with higher weight whereas faraway pixels with lower weight. Thus, Eqn.~\eqref{eqn:merge} assures that the pixel $\bs{x}_{k}$ of the denoised image is heavily influenced by the pixels at the same location of the nearby patches. This single-channel denoising algorithm is summarized in Alg.~\ref{alg:ggd}. This single-channel image denoising scheme is finally extended to a three-channel framework in Sec.~\ref{sec:3ch}.

\begin{algorithm}[H]
\caption{ 
Single-channel denoising \citep{gajamannage2020patch}.\\
Inputs:  noise-contaminated single-channel image ($\mathcal{U}_{n\times n}$), length of a square patch ($\rho$),  size of the nearest neighborhood ($\delta$), and threshold for the SVs ($L$).\\
Outputs:  denoised image ($\tilde{\mathcal{U}}_{n\times n}$).}
\begin{algorithmic}[1]
\State  Produce a set of $n^2$ overlapping $\rho \times \rho$ patches of the noisy image $\mathcal{U}_{n\times n}$ and denote this set by $\{\bs{u}(\bs{x}_k)\vert k=1,\dots,n^2\}$.	
\State Utilize the nearest neighbor search algorithm in \cite{agarwal1999geometric} to construct the graph $G(V,E)$ of the patch set. Produce the geodesic distance matrix $\mathcal{D}$ by approximating the geodesic distances on this graph using Floyd's algorithm in \cite{floyd1962algorithm}.	
\State Use Eqn.~\eqref{eqn:gram} to construct the Gramian matrix $\mathcal{G}$ from $\mathcal{D}$.	
\State Approximate right SVs $\left\{\mathbf{v}_l\vert l = 1,\dots L\right\}$ of the $L$ biggest singular values of the Gramian matrix $\mathcal{G}$ using RSVD in Alg.~\ref{alg:rsvd}.
\State Construct the denoised patches $\{\tilde{\bs{u}}(\bs{x}_k)\vert k=1,\dots,n^2\}$ using the right SVs $\left\{\mathbf{v}_l\vert l = 1,\dots L\right\}$ according to Eqn.~\eqref{eqn:denoise}.	
\State Use Eqns.~\eqref{eqn:merge} and \eqref{eqn:shepard_weight} to merge denoised patches and construct the noise-free image $\tilde{\mathcal{U}}_{n\times n}$.
\end{algorithmic}\label{alg:ggd}
\end{algorithm}

\subsection{Three-channel extension of the single-channel denoising framework}\label{sec:3ch}
Color subsampling takes advantage of the fact that the human visual system has a lot of light-sensitive elements that detect brightness while it has a small number of color-sensitive elements to detect color. Thus, image denoising should be given different focuses for brightness and colors. Here, Luminance, denoted as Y, refers to the brightness of the image, and both Chrominance-blue, denoted as Cb, and Chrominance-red, denoted as Cr, refer to the color of the image. Thus, three-channel RGB images are denoised in three steps: 1) conversion of RGB image into three-channel Luminance, Chrominance-blue, and Chrominance-red (YCbCr) color spectrum using Eqn.~\eqref{eqn:rgb-lc}; 2) denoising each channel of YCbCr using single-channel EGGD; and 3) conversion of denoised YCbCr image back into RGB spectrum using Eqn.~\eqref{eqn:lc-rgb}. 

The transformation of the RGB image into YCrCb is  
\begin{equation}\label{eqn:rgb-lc}
\begin{bmatrix}\textrm{Y}\\ \textrm{Cb}\\ \textrm{Cr}\end{bmatrix} = 
\begin{bmatrix}0\\ 128\\ 128\end{bmatrix}+
\begin{bmatrix}
    0.299 & 0.587 & 0.114 \\
    -0.169 & -0.331 & 0.500 \\    
    0.500 & -0.419 & -0.081 
\end{bmatrix}
\cdot
\begin{bmatrix}\textrm{R}\\ \textrm{G}\\ \textrm{B}\end{bmatrix},
\end{equation}
where, the ranges of all the channels R, G, B, Y, Cr, and Cb are [0, 255]. The transformation of the YCrCb image into RGB is
\begin{equation}\label{eqn:lc-rgb}
\begin{bmatrix}\textrm{R}\\ \textrm{G}\\ \textrm{B}\end{bmatrix} = 
\begin{bmatrix}
   1.000 & 0.000 & 1.400 \\
   1.000 & -0.343 & -0.711 \\    
   1.000 & 1.765 & 0.000 
\end{bmatrix}
\cdot
\begin{bmatrix}\textrm{Y}\\ \textrm{Cb}-128\\ \textrm{Cr}-128\end{bmatrix}.
\end{equation}
The three-channel extension of the denoising algorithm is presented in Alg.~\ref{alg:eggd}. As we will analyze the quality of each channel, R, G, B, Y, Cr, and Cb of the decompositions in Eqns.~\eqref{eqn:rgb-lc} and \eqref{eqn:lc-rgb} with three image quality metrics Shannon Entropy, Peak Signal to Noise Ratio, and Structural Similarity Index Measure; in Sec.~\ref{sec:simMet}, we present the technical details of those metrics and then proceed with analyzing an image.

\begin{algorithm}[H]
\caption{ 
Three-channel denoising with Efficient Geodesic Gramian Denoising (EGGD).\\
Inputs:  noise-contaminated RGB image ($\mathcal{U}_{n\times n\times 3}$), length triplet of square patches [($\rho_1, \rho_2, \rho_3$)],  size triplet of the nearest neighborhoods [($\delta_1, \delta_2, \delta_3$)], and threshold triplet for the SVs [($L_1, L_2, L_3$)].\\
Outputs:  denoised image ($\tilde{\mathcal{U}}_{n\times n\times 3}$).}
\begin{algorithmic}[1]
\State Transform the RGB image $\mathcal{U}_{n\times n\times 3}$  into YCbCr representation using Eqn.~\eqref{eqn:rgb-lc}.	
\State Decompose YCbCr into individual channels Y, Cb, and Cr.
\State Denoise the channels Y, Cb, and Cr separately using Alg.\ref{alg:ggd} with parameters ($\rho, \delta, L$)= ($\rho_1, \delta_1, L_1$), ($\rho_2, \delta_2, L_2$), and ($\rho_3, \delta_3, L_3$), respectively. Let, denoised versions of  Y, Cb, and Cr, and  $\tilde{\textrm{Y}}$, $\tilde{\textrm{Cb}}$, and $\tilde{\textrm{Cr}}$, respectively.
\State Transform $\tilde{\textrm{Y}}$, $\tilde{\textrm{Cb}}$, and $\tilde{\textrm{Cr}}$ using  Eqn.~\eqref{eqn:lc-rgb} to get the denoised version $\tilde{\mathcal{U}}_{n\times n\times 3}$ of the noisy input image.
\end{algorithmic}\label{alg:eggd}
\end{algorithm}

\section{Performance Analysis}\label{sec:per}
Here, first, we present image similarity metrics that we utilize to analyze the performance of remote sensing image denoising; then, we present real-life image denoising examples.

\subsection{Similarity Metrics}\label{sec:simMet}
We employ three benchmark similarity metrics, Shannon Entropy, Peak Signal to Noise Ratio, and Structural Similarity Index Measure, to analyze the performance.

\subsubsection{Shannon Entropy}
Shannon Entropy, denoted as ShE and defined in Def.~\ref{def:she}, is a measure of the average information content of an image \citep{shannon1959coding}. In Shannon's information theory, entropy is a measure of the uncertainty over the true content of an image, but the task is complicated by the fact that successive bits in an image are not random, and therefore not mutually independent in a real image \citep{cover1999elements}. ShE of an 8-bit image ranges between 0 and 8 where 0 represents a certain image with less variability while 8 represents a highly uncertain image. 
\vspace{2mm}
\begin{definition}\label{def:she}
Let, $p_i$ be the probability of occurrence of a given pixel where a single-channel 8-bit image has occurrences from 0 to 255. Shannon Entropy \citep{shannon1959coding}, abbreviated as ShE, of the image $\tilde{\mathcal{U}}$ is defined as  
\begin{equation}
ShE(\tilde{\mathcal{U}}) = - \sum_{i=1}^n p_i \log_2(p_i).
\end{equation}
Here, $n$ is the maximum possible occurrences for an arbitrary image pixel.
\end{definition}
\vspace{2mm}

\subsubsection{Peak Signal to Noise Ratio}
Peak Signal to Noise Ratio, denoted as PSNR and defined in Def.~\ref{def:psnr}, measures the numerical difference of an image from a reference image, with respect to the maximum possible pixel value of the reference image \citep{hore2010image} where PSNR ranges between $0$ and $\infty$. If the original noise-free image is the reference image and a denoised approximation of its noisy version is the other image, a higher PSNR value provides a better quality of the approximation since it approaches infinity as the Root Mean Square Error (RMSE), see Def.~\ref{def:rmse}, approaches zero \citep{hore2010image}. 
\begin{definition}\label{def:rmse}
Let, two-dimensional matrix $\mathcal{I}$ represents a reference image of size $n\times n$ and $\tilde{\mathcal{U}}$ represents any other image of interest. Root Mean Square Error \citep{hore2010image}, abbreviated as RMSE, of the image $\tilde{\mathcal{U}}$ with respect to the reference image $\mathcal{I}$ is defined as  
\begin{equation}
RMSE(\mathcal{I},\tilde{\mathcal{U}}) = \sqrt{\frac{\sum_{(i,j)\in\mathbb{N}^{n\times n}} (\mathcal{I}[i,j]-\tilde{\mathcal{U}}[i,j])^2}{n^2}}.
\end{equation}
\end{definition}   
\vspace{2mm}
\begin{definition}\label{def:psnr}
Let, two-dimensional matrix $\mathcal{I}$ represents a reference image of size $n\times n$ and $\tilde{\mathcal{U}}$ represents any other image of interest. Peak Signal to Noise Ratio \citep{hore2010image}, abbreviated as PSNR, of the image $\tilde{\mathcal{U}}$ with respect to the reference image $\mathcal{I}$ is defined as  
\begin{equation}
PSNR(\mathcal{I},\tilde{\mathcal{U}}) = 20 \log_{10}\left(\frac{\max(\mathcal{I})}{RMSE(\mathcal{I},\tilde{\mathcal{U}})}\right).
\end{equation}
Here, $\max(\mathcal{I})$ represents the maximum possible pixel value of the image $\mathcal{I}$. Since the pixels in our images of interest are represented in 8-bit digits, $\max(\mathcal{I})$ is 255.
\end{definition}   
\vspace{2mm}

\subsubsection{Structural Similarity Index Measure}
The Structural Similarity Index Measure, denoted as SSIM and defined in Def.~\ref{def:ssim}, ranging between -1 and 1 measures the structural similarity of an image to a reference image \citep{wang2004image}. Ideal approximation provides 1 for SSIM since such approximation ensures $\mu_{\mathcal{I}}=\mu_{\tilde{\mathcal{U}}}$, $\sigma_{\mathcal{I}}=\sigma_{\tilde{\mathcal{U}}}$, and $\sigma_{\mathcal{I}\tilde{\mathcal{U}}}=\sigma_{\mathcal{I}}\sigma_{\tilde{\mathcal{U}}}$ in Def.~\ref{def:ssim}. Instead of using traditional error summation methods, SSIM is designed by modeling any image distortion as a combination of three factors, namely, luminance distortion, contrast distortion, and loss of correlation. 
\begin{definition}\label{def:ssim}
Let, two-dimensional matrix $\mathcal{I}$ represents a reference image of size $n\times n$ and $\tilde{\mathcal{U}}$ represents an image of interest. Structural Similarity Index Measure \citep{wang2004image}, abbreviated as SSIM, of the image $\tilde{\mathcal{U}}$ with respect to the reference image $\mathcal{I}$ is defined as the product of luminance distortion ($I$), contrast distortion ($C$), and loss of correlation ($S$), such as
\begin{equation}
SSIM(\mathcal{I},\tilde{\mathcal{U}}) = I(\mathcal{I},\tilde{\mathcal{U}}) \ C(\mathcal{I},\tilde{\mathcal{U}}) \ S(\mathcal{I},\tilde{\mathcal{U}}),
\end{equation}
where
\begin{equation}
\begin{split}
I(\mathcal{I},\tilde{\mathcal{U}}) = \frac{2\mu_{\mathcal{I}}\mu_{\tilde{\mathcal{U}}}+c_1}{\mu^2_{\mathcal{I}}+\mu^2_{\tilde{\mathcal{U}}}+c_1},\\
C(\mathcal{I},\tilde{\mathcal{U}}) = \frac{2\sigma_{\mathcal{I}}\sigma_{\tilde{\mathcal{U}}}+c_2}{\sigma^2_{\mathcal{I}}+\sigma^2_{\tilde{\mathcal{U}}}+c_2},\\
S(\mathcal{I},\tilde{\mathcal{U}}) = \frac{\sigma_{\mathcal{I}\tilde{\mathcal{U}}}+c_3}{\sigma_{\mathcal{I}}\sigma_{\tilde{\mathcal{U}}}+c_3}.\\
\end{split}
\end{equation}
Here, $\mu_{\mathcal{I}}$ and $\mu_{\tilde{\mathcal{U}}}$ are means of $\mathcal{I}$ and $\tilde{\mathcal{U}}$, respectively; $\sigma_{\mathcal{I}}$ and $\sigma_{\tilde{\mathcal{U}}}$ are standard deviations of $\mathcal{I}$ and $\tilde{\mathcal{U}}$, respectively; and  $\sigma_{\mathcal{I}\tilde{\mathcal{U}}}$ is the covariance between $\mathcal{I}$ and $\tilde{\mathcal{U}}$. Moreover, $c_1$, $c_2$, and $c_3$ are very small positive constants to avoid the case of division by zero.
\end{definition}   
\vspace{2mm}

\begin{figure*}[htp]
\centering
\includegraphics[width=1\textwidth]{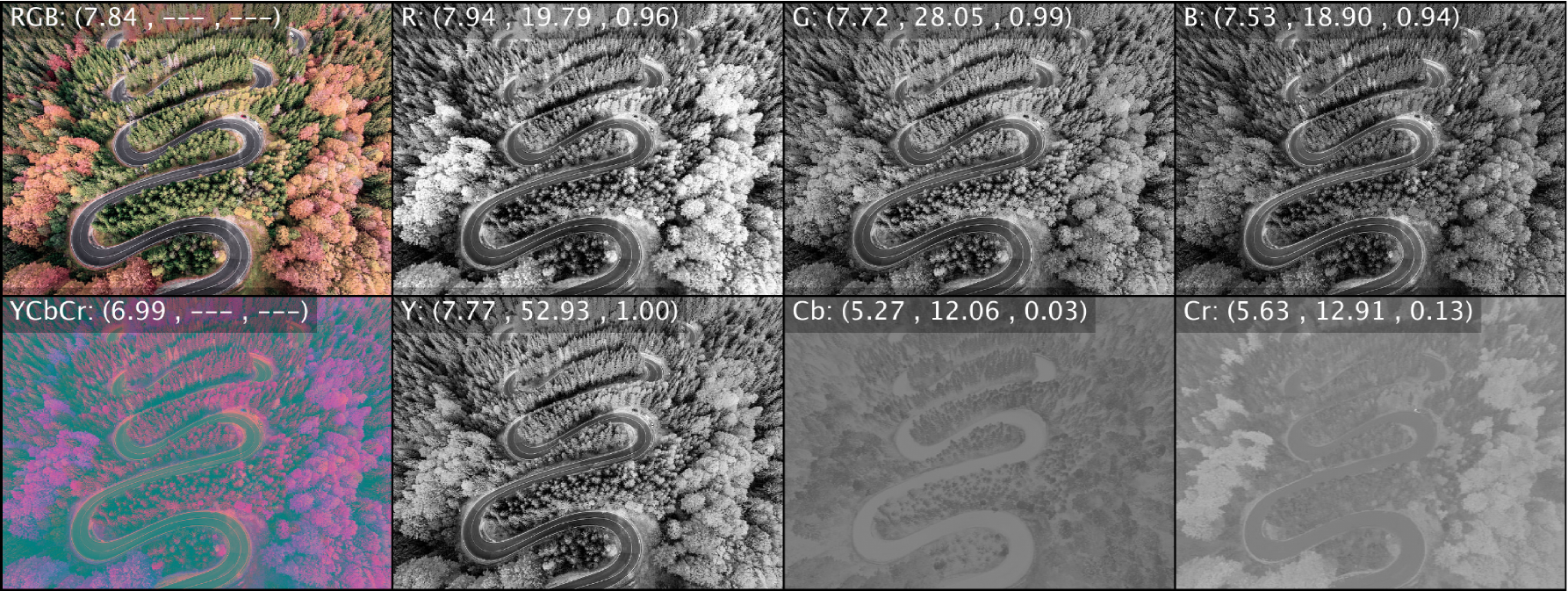}
\caption{A red-green-blue image}, denoted as RGB, is decomposed into its color channels, Red (R), Green (G), and Blue (B). The luminance-chrominance transformation, denoted as YCbCr, of the RGB image is decomposed into its individual color channels  Luminance (Y), Chrominance-blue (Cb), and Chrominance-red (Rb). The triplet $(a, b, c)$ of each image represents its Shannon entropy by $a$, Point-Signal-to-Noise-Ratio (PSNR) by $b$, and Structural-Similarity-Index-Measure (SSIM) by $c$. The image RGB is used as the reference image for the computation of PSNR and SSIM of the images R, G, and B, and the image YCbCr is used as the reference image for the computation of PSNR and SSIM of the images Y, Cb, and Cr; thus, the second and the third values of the triplets are empty in RGB and YCbCr.
\label{fig:decom}
\end{figure*}

Image decomposition and conversion that are presented in Sec.~\ref{sec:3ch} is analyzed using an arbitrary remote sensing image of vegetation. The image, denoted as RGB in Fig.~\ref{fig:decom}, is decomposed into its three channels, red, green, and blue, represented in Fig.~\ref{fig:decom} as R, G, and B, respectively. The RGB image is transformed into its luminance-chrominance color spectrum, denoted as YCbCr in Fig.~\ref{fig:decom}, using Eqn.~\eqref{eqn:rgb-lc}.  YCbCr image is decomposed into three channels, luminance, chrominance-blue, and chrominance-red, represented in Fig.~\ref{fig:decom} as Y, Cb, and Cr, respectively. For each of the eight images in Fig.~\ref{fig:decom}, we compute ShE using Def.~\ref{def:she}. For the images R, G, and B, we compute PSNR using Def.~\ref{def:psnr} and SSIM using Def.~\ref{def:ssim}, with the reference image RGB. For the images Y, Cb, and Cr, we compute PSNR and SSIM with the reference image YCbCr. The triplet $(a, b, c)$ presents ShE, PSNR, and SSIM in that order, where the images RGB and YCbCr don't have PSNR and SSIM values as they are reference images. 

We observe in Fig.~\ref{fig:decom} that ShE, PSNR, and SSIM values for the images R, G, and B are similar. However, ShE, PSNR, and SSIM values for the images Cb and Cr are considerably lower than those of the image Y. Thus, channel Y of the YCbCr representation possesses the greatest information (measured by ShE) while being similar to the original RGB image in both pixel values (measured by PSNR) and structure (measured by SSIM). As seen in this example, denoising RGB images directly through channels R, G, and B would not be efficient since it increases computational cost as each channel should be treated with the same emphasis. Moreover, minor changes in the pixel intensities on channels R, G, and B during the denoising process might change the features of the RGB image significantly as each channel influences the original image equally. Conversely, the YCbCr representation of an RGB image only segregates features into Y while Cb and Cr retain blue and red color residues, respectively. Thus, EGGD transforms an RGB image into the YCbCr representation using Eqn.~\eqref{eqn:rgb-lc} and then asserts more emphasis only on the Y channel during denoising. However, since remote sensing images contain more information than regular images, better emphasis should be given to both Cb and Cr channels for the superior quality of image denoising. EGGD insists primary focus on Y while imposing adequate emphasis on the other two channels Cb and Cr. The adapted emphasis of the denoising strength between the channels Y, Cb, and Cr is attained by utilizing diverse values for the parameters $\rho$, $\delta$, and $L$ for EGGD across channels as we explained in Sec.~\ref{sec:exa}. 

\subsection{Remote Sensing Image Denoising Examples}\label{sec:exa}
The performance of the EGGD framework is evaluated on seven diverse remote sensing test images, desert, ocean, snow, volcano, watershed, city, and cargo. Here, first, we distort our test images with Gaussian noise and then attempt to filter the noise out using EGGD. The probability density function (PDF) of Gaussian noise is given by
\begin{equation}\label{eqn:gauss}
P(z,\mu,\gamma) = {\frac{1}{\sigma \sqrt{2 \pi}} { e^{ \frac{ -{(z-\mu)}^2 } { 2 \sigma^2 } } } },
\end{equation}
where $z$ is gray-level, $\mu$ is the mean of the distribution, and $\sigma$ is the standard deviation of the distribution.

Consider that we are given a noise-free RGB image of size $n\times n\times 3$, denoted as $\mathcal{I}_{n\times n\times 3}$. We draw a noise sample of size $n\times n\times 3$, denoted as $\mathcal{N}_{n\times n\times 3}$, from the Gaussian PDF given in Eqn.~\eqref{eqn:gauss} with $\mu=0$ and some $\sigma$. Noisy image, denoted as $\mathcal{U}_{n\times n\times 3}$, is generated by additive rule as
\begin{equation}
\mathcal{U}_{n\times n\times 3}=\mathcal{I}_{n\times n\times 3}+\mathcal{N}_{n\times n\times 3}.
\end{equation}
We compute the relative percentage noise, denoted as $\zeta$, by 
\begin{equation}\label{eqn:noise}
\zeta=\frac{\|\mathcal{U}-\mathcal{I}\|_2}{\|\mathcal{I}\|_2} 100\%.
\end{equation}
The original noise-free image $\mathcal{I}_{10^3 \times 10^3\times 3}$ is imposed with diverse levels of noise intensities by changing $\sigma$ of the PDF in Eqn.~\eqref{eqn:gauss} to create three noisy images with $\zeta=2\%$, $4\%$, and $6\%$. Since $\mathcal{U}$'s represent images, each entry of these matrices should be between 0--255 even after imposing the noise. Thus, we adjust $\mathcal{U}$ by replacing the values less than zero with zeros and the values more than 255 with 255s. 

The noisy test images are denoised with EGGD as well as four state-of-the-art methods which two of them, BM3D and KSVD, are linear algebraic techniques, and the other two, ADNet and DnCNN, are ANN techniques. For the test images with $\zeta=2\%$, we implement EGGD with the parameter triplets ($\rho, \delta, L$) = (5, 20, 80), (7, 20, 80), and (7, 20, 80) for the channels Y, Cr, and Cb, respectively. We arbitrarily choose these parameter values except for setting small values like 5 and 7 for the patch size $\rho$ since the noise contamination here is minimal. We consistently set $\delta=20$ and $L=80$ for all the channels and the experiments in this paper. Each BM3D and KSVD possesses a parameter inferring the noise contamination of the images; thus, we use $\epsilon$ to denote the standard deviation of the Gaussian noise contamination of noisy images. We implement BM3D with its parameters denoising strength (measures as the standard deviation of the imposed noise), patch size, sliding step size, the maximum number of similar blocks, radius for search block matching, step between two search locations, 2D thresholding, 3D thresholding, and threshold for the block-distance are set to the values $\epsilon$, 12, 4, 16, 39, 1, 2, 2.8, and 3000, respectively, as recommended by its literature. We implement KSVD with its parameters block size, dictionary size, number of training iterations, Lagrangian multiplier, noise gain, and number of non-zero coefficients are set to the values 64, 244, 10, 30/$\epsilon$, 1.55, and 2, respectively, as recommended by its literature. Here, we explicitly find the corresponding $\epsilon$ for the relative noise level $\zeta=2\%$ and then use that for BM3D and KSVD. We use the pre-trained ANNs of ADNet and DnCNN given in their literature, and denoised test images with 17 layers in ADNet and 20 layers in DnCNN as recommended by their literature. 

\begin{figure*}[htp]
\centering
\includegraphics[width=.95\textwidth]{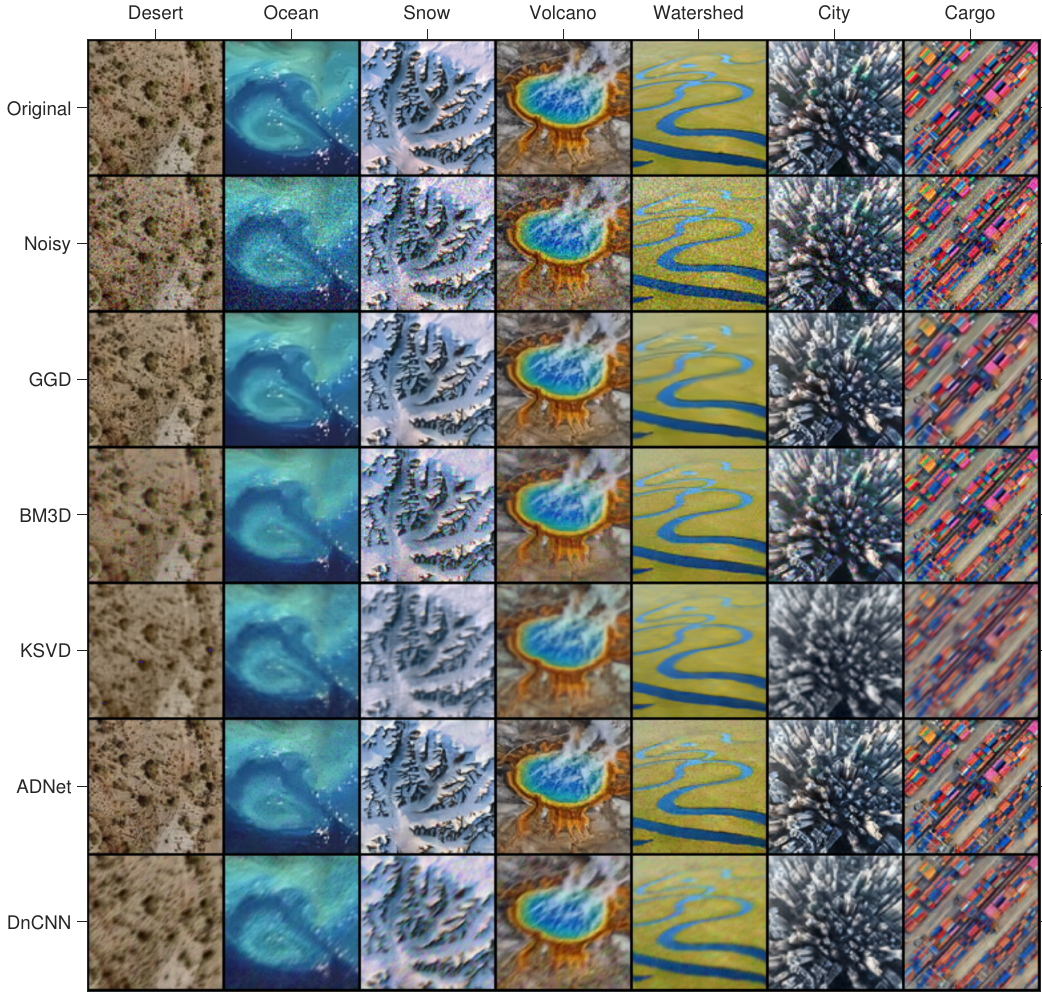}
\caption{Remote sensing image denoising with EGGD and four other state-of-the-art denoising methods, namely, sparse 3-D transform-domain collaborative filtering (BM3D), sparse and redundant representations over learned dictionaries (KSVD), Attention-guided Denoising Convolutional Neural Network (ADNet), and Denoising Convolutional Neural Networks (DnCNN). Each of the seven test images, namely, desert, ocean, snow, volcano, watershed, city, and cargo, of size $10^3 \times 10^3$, is imposed with a relative Gaussian noise of 2\%. EGGD is executed with the parameter values ($\rho, \delta, L$) = (5, 20, 80), (7, 20, 80), and (7, 20, 80) for the channels Y, Cr, and Cb, respectively. The other four methods are executed using their recommended parameter values, or tested using pre-train ANNs given in their literature. Here, we observe that EGGD retains texture and cartoon in the denoised images most of the time than that of the other four methods.}
\label{fig:compNoiL1}
\end{figure*}

\begin{figure*}[htp]
\centering
\includegraphics[width=1\textwidth]{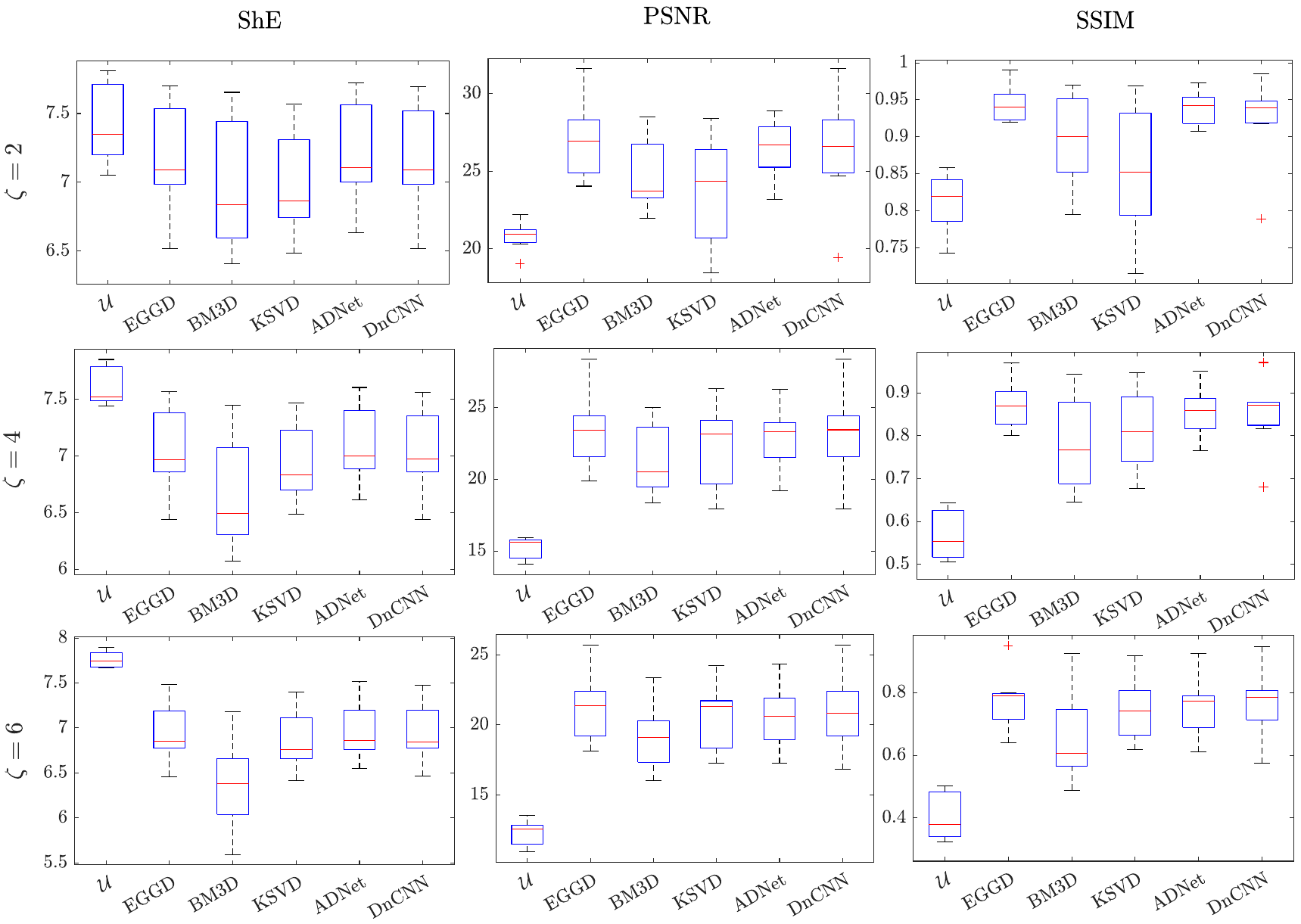}
\caption{Box and whisker plots of the three similarity metrics, Shannon Entropy, denoted by ShE, Peak Signal to Noise Ratio, denoted by PSNR, and Structural Similarity Index Measure, denoted by SSIM, of the experiments of remote sensing image denoising with EGGD and four other state-of-the-art denoising methods, namely, sparse 3-D transform-domain collaborative filtering (BM3D), sparse and redundant representations over learned dictionaries (KSVD), Attention-guided Denoising Convolutional Neural Network (ADNet), and Denoising Convolutional Neural Networks (DnCNN). Seven remote sensing test images each of size $10^3 \times 10^3$ are imposed with three relative Gaussian noise levels, denoted as $\zeta$, of $2\%$, $4\%$, and $6\%$ to make noisy images, denoted by $\mathcal{U}$. For $\zeta=2\%$, EGGD is implemented with the parameter values ($\rho, \delta, L$) = (5, 20, 80), (7, 20, 80), and (7, 20, 80) for the channels Y, Cr, and Cb, respectively. For $\zeta=4\%$, EGGD is implemented with the parameter values ($\rho, \delta, L$) = (7, 20, 80), (9, 20, 80), and (9, 20, 80) for the channels Y, Cr, and Cb, respectively. For $\zeta=6\%$, EGGD is implemented with the parameter values ($\rho, \delta, L$) = (9, 20, 80), (11, 20, 80), and (11, 20, 80) for the channels Y, Cr, and Cb, respectively. For a given noise level and a method, the seven denoised images, desert, ocean, snow, volcano, watershed, city, and cargo are considered to be seven trials. Thus, we compute box and whisker plots of the values of a denoising performance metric of interest (i.e., ShE, PSNR, and SSIM) over these trials. While BM3D and KSVD are implemented using the recommended parameter values mentioned in their literature, ADNet and DnCNN are implemented using pre-train ANNs given in their literature.}
\label{fig:boxPlot}
\end{figure*}

Fig.~\ref{fig:compNoiL1} represents the original images, noisy images, and denoised images for the noise level $\zeta=2\%$. We visually observe that EGGD and ADNet perform better than the other three methods. To numerically assess this observation, we compute, 1)  three similarity metrics, ShE, PSNR, and SSIM for each image denoised by EGGD and four other state-of-the-art denoising which we show in Table~\ref{tab}; and 2) box and whisker plots for the similarity metrics of denoising with each method which we show in Fig.~\ref{fig:boxPlot}. For a given noise level and a method, the seven denoised images, desert, ocean, snow, volcano, watershed, city, and cargo are considered to be seven trials. Thus, we compute mean, standard deviations, and box and whisker plots, of the values of a denoising performance metric of interest (i.e., ShE, PSNR, and SSIM) over these trials. Average ShE indicates that the information content of the original image is closer to that of denoised images of EGGD and ADNet than the other three methods. The largest average PSNR and SSIM values are attained by both EGGD and ADNet; thus, both EGGD and ADNet possess the highest numerical and structural similarity to the corresponding original image than the other two methods. 

\begin{table*}[htp]
\centering
\caption{Comparison of the remote sensing image denoising performance of EGGD and four benchmark denoising methods, namely, sparse 3-D transform-domain collaborative filtering (BM3D), sparse and redundant representations over learned dictionaries (KSVD), Attention-guided Denoising Convolutional Neural Network (ADNet), and Denoising Convolutional Neural Networks (DnCNN). For that, seven remote sensing test images, namely, desert, ocean, snow, volcano, watershed, city, and cargo, each of size $10^3 \times 10^3$, denoted by $\mathcal{I}$, are imposed with three relative Gaussian noise levels, denoted as $\zeta$, of $2\%$, $4\%$, and $6\%$ to make noisy images, denoted by $\mathcal{U}$. The comparison is performed with respect to three similarity metrics, Shannon Entropy, denoted by ShE (the first value in each cell), Peak Signal to Noise Ratio, denoted by PSNR, (the second value in each cell), and Structural Similarity Index Measure, denoted by SSIM (the third value in each cell). ShE of the original images and all three metrics of the noisy images are also computed. This table shows the mean and standard deviation (denoted as SD and shown in parenthesis) of each similarity metric over the test images for each noise level, and also it shows the overall mean and SD. Here, one trial is considered to be one denoised image for a given $\zeta$ and denoising method. For $\zeta=2\%$, EGGD is implemented with the parameter values ($\rho, \delta, L$) = (5, 20, 80), (7, 20, 80), and (7, 20, 80) for the channels Y, Cr, and Cb, respectively. For $\zeta=4\%$, EGGD is implemented with the parameter values ($\rho, \delta, L$) = (7, 20, 80), (9, 20, 80), and (9, 20, 80) for the channels Y, Cr, and Cb, respectively. For $\zeta=6\%$, EGGD is implemented with the parameter values ($\rho, \delta, L$) = (9, 20, 80), (11, 20, 80), and (11, 20, 80) for the channels Y, Cr, and Cb, respectively. While BM3D and KSVD are implemented using the recommended parameter values mentioned in their literature, ADNet and DnCNN are implemented using pre-train ANNs given in their literature.}
\footnotesize
\begin{tabular}{| p{.3 cm} | p{1.6 cm} | p{.8 cm} | p{1 cm} | p{1.2 cm} | p{1.2 cm} | p{1.2 cm} | p{1.2cm} | p{1.2 cm} |}
\hline
\multicolumn{2}{|c|}{} & $\mathcal{I}$ & $\mathcal{U}$ & EGGD & BM3D & KSVD & ADNet & DnCNN \\
\hline
\hline
\multirow{24}{*}{\rotatebox[origin=c]{90}{Images with $\zeta = 2$}} & \multirow{3}{*}{Desert}
     & 7.04 & 7.17 & 6.98 & 6.55 & 6.17 & 6.99 & 6.83 \\
 & & ----- & 22.20 & 28.47 & 23.18 & 24.39 & 28.07 & 23.27 \\
 & & ----- & 0.82 & 0.95 & 0.79 & 0.85 & 0.94 & 0.81 \\
\cline{2-9}	
 & \multirow{3}{*}{Ocean}
    & 7.05 & 7.35 & 7.00 & 6.75 & 6.98 & 7.02 & 7.00 \\
 & & ----- & 20.66 & 31.62 & 27.48 & 28.38 & 28.8 & 27.43 \\
 & & ----- & 0.82 & 0.99 & 0.96 & 0.97  & 0.97 & 0.96 \\
\cline{2-9}
 & \multirow{3}{*}{Snow}
     & 7.70 & 7.78 & 7.62 & 7.51 & 7.42 & 7.65 & 7.45 \\
 & & ----- & 19.03 & 24.72 & 21.9 & 20.50 & 24.84 & 19.99 \\
 & & ----- &  0.74 & 0.92 & 0.85 & 0.79 & 0.91 & 0.77 \\
\cline{2-9}
 & \multirow{3}{*}{Volcano}
    & 7.16 & 7.29 & 7.09 & 6.84 & 6.85 & 7.11 & 6.99 \\
 & & ----- & 21.08 & 27.88 & 24.56 & 24.34 & 27.26 & 24.25 \\
 & & ----- &  0.78 & 0.94 & 0.85 & 0.86 & 0.93 & 0.85 \\
\cline{2-9}
 & \multirow{3}{*}{Watershed}
     & 6.62 & 7.05 & 6.52 & 6.41 & 6.49 & 6.63 & 6.63 \\
 & & ----- & 20.30 & 25.35 & 28.50 & 27.05 & 26.72 & 25.36 \\
 & & ----- &  0.83 & 0.92 & 0.97 & 0.96 & 0.96 & 0.93 \\
\cline{2-9}
 & \multirow{3}{*}{City}
     & 7.75 & 7.81 & 7.70 & 7.65 & 7.57 & 7.72 & 7.69 \\
 & & ----- & 21.30 & 26.92 & 23.74 & 21.22 & 26.51 & 26.59 \\
 & & ----- & 0.85 & 0.96 & 0.90 & 0.81 & 0.95 & 0.95 \\
\cline{2-9}
& \multirow{3}{*}{Cargo}
     & 7.41 & 7.49 & 7.28 & 7.25 & 6.87 & 7.31 & 7.21 \\
 & & ----- & 20.95 & 24.04 & 23.56 & 18.46 & 23.20 & 19.45 \\
 & & ----- & 0.86 & 0.93 & 0.91 & 0.72 & 0.91 & 0.79 \\
\cline{2-9}
& \multirow{3}{*}{Mean (SD)}
     & 7.25 (0.37) & 7.42 (0.27) & 7.17 (0.38) & 6.99 (0.45) & 6.98 (0.36) & 7.21 (0.36) & 7.16 (0.37) \\
 & & ----- & 20.79 (0.90) & 26.98 (2.41) & 24.71 (2.21) & 23.48 (3.33) & 26.50 (1.79)& 26.30 (3.50)\\
 & & ----- & 0.81 (0.04) & 0.94 (0.02) & 0.89 (0.06) & 0.85 (0.08) & 0.94 (0.02) & 0.92 (0.06) \\
 \hline
\hline
\multirow{24}{*}{\rotatebox[origin=c]{90}{Images with $\zeta = 4$}} & \multirow{3}{*}{Desert} 
    & 7.04 & 7.44 & 6.86 & 6.07 & 6.66 & 6.89 & 6.77 \\
 & & ----- & 15.70 & 23.98 & 20.50 & 23.24 & 23.68 & 21.76 \\
 & & ----- &  0.51 & 0.85 & 0.65	& 0.81 & 0.84 & 0.74 \\
\cline{2-9}
 & \multirow{3}{*}{Ocean} 
     & 7.05 & 7.52 & 6.86 & 6.50 & 6.94 & 6.90 & 6.96 \\
 & & ----- & 15.63 & 28.37 & 24.98 & 26.30 & 26.25 & 25.39 \\
 & & ----- &  0.60 & 0.97 & 0.94 & 0.95 & 0.95 & 0.94 \\
\cline{2-9}
 & \multirow{3}{*}{Snow} 
    & 7.70 & 7.85 & 7.44 & 7.07 & 7.33 & 7.48 & 7.55 \\
 & & ----- & 14.08 & 21.18 & 18.37 & 19.43 & 21.16 & 18.37 \\
 & & ----- &  0.51 & 0.82 & 0.67	& 0.73 & 0.81 & 0.66  \\
\cline{2-9}
 & \multirow{3}{*}{Volcano} 
    & 7.16 & 7.48 & 6.97 & 6.45 & 6.82 & 7.00 & 6.92 \\
 & & ----- & 15.82 & 24.55 & 21.45 & 23.14 & 24.01 & 22.27 \\
 & & ----- &  0.54 & 0.87 & 0.74 & 0.82 & 0.86 & 0.78 \\
\cline{2-9}
 & \multirow{3}{*}{Watershed}
    & 6.62 & 7.51 & 6.44 & 6.26 & 6.49 & 6.62 & 6.53 \\
 & & ----- & 14.24 & 23.43 & 24.34 & 24.36 & 23.28 & 23.06 \\
 & & ----- &  0.55 & 0.88 & 0.91 & 0.92 & 0.89 & 0.88  \\
\cline{2-9}
 & \multirow{3}{*}{City}
     & 7.75 & 7.84 & 7.57 & 7.45 & 7.47 & 7.61 & 7.56 \\
 & & ----- & 15.92 & 22.68 & 20.03 & 20.39 & 22.60 & 22.77 \\
 & & ----- & 0.64 & 0.92 & 0.77 & 0.78 & 0.87 & 0.88 \\
\cline{2-9}
 & \multirow{3}{*}{Cargo}
     & 7.41 & 7.64 & 7.20 & 7.07 & 6.83 & 7.18 & 7.11 \\
 & & ----- & 15.22 & 19.88 & 19.24 & 17.91 &19.20 & 17.91 \\
 & & ----- & 0.64 & 0.80 & 0.77 & 0.68 & 0.77 & 0.68 \\
\cline{2-9}
 & \multirow{3}{*}{Mean (SD)}
     & 7.25 (0.37) & 7.61 (0.16) & 7.05 (0.36) & 6.70 (0.47) & 6.93 (0.33) & 7.10 (0.32) & 7.03 (0.35)\\
 & & ----- & 15.23 (0.71) & 23.44 (2.51) & 21.27 (2.33) & 22.11 (2.75) & 22.88 (2.07) & 23.17 (2.96)\\
 & & ----- &  0.57 (0.05) & 0.87 (0.05) & 0.78 (0.10) & 0.81 (0.09) & 0.86 (0.06) & 0.85 (0.08) \\
\hline
\end{tabular}

\label{tab}
\end{table*}

\begin{table*}[htp]
\centering
\footnotesize
\begin{tabular}{| p{.3 cm} | p{1.6 cm} | p{.8 cm} | p{1 cm} | p{1.2 cm} | p{1.2 cm} | p{1.2 cm} | p{1.2cm} | p{1.2 cm} |}
\hline
\multirow{24}{*}{\rotatebox[origin=c]{90}{Images with $\zeta = 6$}} & \multirow{3}{*}{Desert} 
    & 7.04 & 7.67 & 6.76 & 5.59 & 6.63 & 6.72 & 6.69 \\
 & & ----- & 12.53 & 22.01 & 19.08 & 21.73 & 21.65 & 20.45 \\
 & & ----- &  0.32&  0.76 & 0.55 & 0.74 & 0.73 & 0.65  \\			
\cline{2-9}
 & \multirow{3}{*}{Ocean} 
    & 7.05 & 7.67 & 6.82 & 6.38 & 6.92 & 6.85 & 6.86 \\
 & & ----- & 12.84 & 25.69 & 23.40 & 24.25 & 24.32 & 24.02 \\
 & & ----- &  0.43 & 0.95 & 0.92 & 0.92 & 0.92 & 0.91 \\
\cline{2-9}
 & \multirow{3}{*}{Snow} 
    & 7.70 & 7.89 & 7.27 & 6.60 & 7.18 & 7.27 & 7.33 \\
 & & ----- & 11.12 & 18.67 & 16.03 & 17.95 & 18.46 & 16.95 \\
 & & ----- &  0.34 & 0.70 & 0.49 & 0.64 & 0.68 & 0.54 \\
\cline{2-9}
 & \multirow{3}{*}{Volcano}
    & 7.16 & 7.68 & 6.85 & 6.07 & 6.75 & 6.86 & 6.71 \\
 & & ----- & 12.80 & 22.54 & 19.73 & 21.70 & 21.97 & 20.63 \\
 & & ----- &  0.38 & 0.80 & 0.64 & 0.76 & 0.77 & 0.69  \\
\cline{2-9}
 & \multirow{3}{*}{Watershed}
    & 6.62 & 7.77 & 6.46 & 6.03 & 6.42 & 6.55 & 6.35 \\
 & & ----- & 10.93 & 20.83 & 20.46 & 21.30 & 20.45 & 20.96\\
 & & ----- &  0.34 & 0.79 & 0.78 & 0.82 & 0.77 & 0.81 \\
\cline{2-9}
 & \multirow{3}{*}{ City}
     &  7.75 &  7.86 &  7.48 &  7.18 & 7.40 &  7.51 &  7.48 \\
 & & ----- &  13.52 &  21.35 &  17.89 &  19.51 &  20.61 & 20.84 \\
 & & ----- & 0.50 & 0.79 & 0.60 & 0.73 & 0.80 &  0.81\\
\cline{2-9}
 & \multirow{3}{*}{Cargo}
     & 7.41 & 7.74 & 6.95 & 6.68 & 6.76 & 6.99 & 6.98 \\
 & & ----- & 12.59 & 18.15 & 17.15 & 17.26 & 17.28 & 16.82 \\
 & & ----- & 0.50 & 0.64 & 0.61 & 0.62 & 0.61 & 0.57 \\
\cline{2-9}
& \multirow{3}{*}{Mean (SD)}
     & 7.25 (0.37) & 7.76 (0.09) & 6.94 (0.31) & 6.36 (0.48) & 6.87 (0.31) & 6.96 (0.30) & 6.95 (0.31)\\
 & & ----- & 12.33 (0.88) & 21.32 (2.34) & 19.10 (2.25) & 20.53 (2.26) & 20.68 (2.15) & 21.06 (2.63)\\
 & & ----- &  0.40 (0.07) &  0.78 (0.09)& 0.66 (0.14) & 0.75 (0.09) & 0.75 (0.09) & 0.77 (0.11) \\
\hline
\hline
\multicolumn{2}{|l|}{\multirow{3}{*}{\begin{tabular}{@{}c@{}}Overall Mean \\ (Overall SD)\end{tabular}}}
& 7.25 (0.37) & 7.60 (0.23) & 7.05 (0.36) & 6.68 (0.53) & 6.93 (0.33) & 7.09 (0.34) & 7.05 (0.36)\\
\multicolumn{2}{|l|}{} & ---- & 16.12 (3.61) & 23.91 (3.37) & 21.70 (3.23) & 22.04 (3.06) & 23.35 (3.13) & 23.51 (3.74)\\
\multicolumn{2}{|l|}{} & ---- & 0.59 (0.18) & 0.86 (0.09) & 0.78 (0.14) & 0.80 (0.10) & 0.85 (0.10) & 0.85 (0.10)\\
\hline
\end{tabular}
\end{table*}

\begin{figure*}[htp]
\centering
\includegraphics[width=1\textwidth]{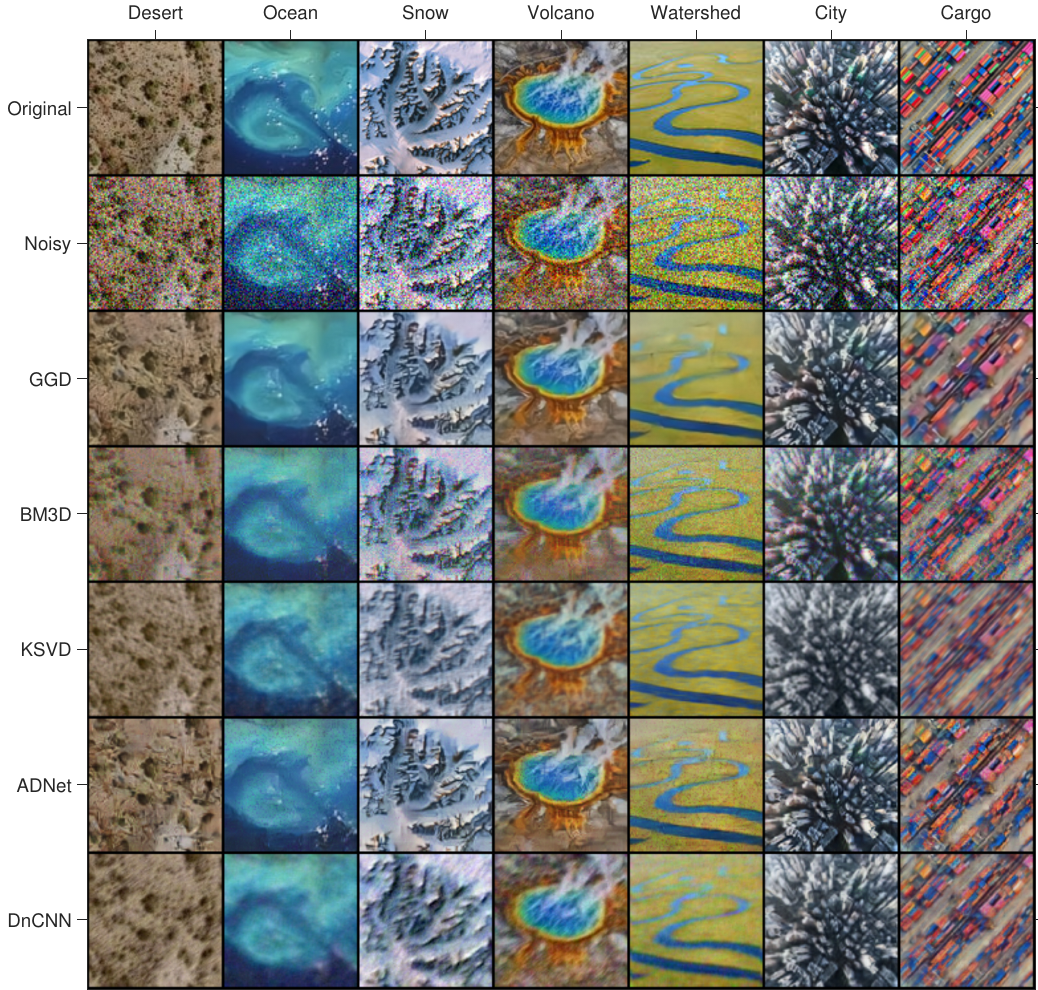}
\caption{Remote sensing image denoising with EGGD and four other state-of-the-art denoising methods, namely, sparse 3-D transform-domain collaborative filtering (BM3D), sparse and redundant representations over learned dictionaries (KSVD), Attention-guided Denoising Convolutional Neural Network (ADNet), and Denoising Convolutional Neural Networks (DnCNN). Each of the seven test images, namely, desert, ocean, snow, volcano, watershed, city, and cargo, of size $10^3 \times 10^3$, is imposed with a relative Gaussian noise of 4\%. EGGD is executed with the parameter values ($\rho, \delta, L$) = (7, 20, 80), (9, 20, 80), and (9, 20, 80) for the channels Y, Cr, and Cb, respectively. The other four methods are executed with their recommended parameter values, or tested with their given trained ANNs. Here, we observe that EGGD retains texture and cartoon in the denoised images most of the time than that of the other four methods.}
\label{fig:compNoiL2}
\end{figure*}

For the noise level $\zeta=4\%$, we implement EGGD with the parameter triplets ($\rho, \delta, L$) = (7, 20, 80), (9, 20, 80), and (9, 20, 80) for the channels Y, Cr, and Cb, respectively. We arbitrarily choose these parameter values except for setting slightly bigger values like 7 and 9 for the patch size $\rho$ than that of $\zeta=2\%$, since the noise contamination in the case $\zeta=4\%$ is bigger. We implement BM3D with its parameters denoising strength, patch size, sliding step size, the maximum number of similar blocks, radius for search block matching, step between two search locations, 2D thresholding, 3D thresholding, and threshold for the block-distance of BM3D are set to $\epsilon$, 12, 4, 16, 39, 1, 2, 2.8, and 3000, respectively. We implement KSVD with its parameters block size, dictionary size, number of training iterations, Lagrangian multiplier, noise gain, and number of non-zero coefficients of KSVD are set to their recommended values 64, 244, 10, 30/$\epsilon$, 1.55, and 2, respectively. Here, we explicitly find the corresponding $\epsilon$ for the relative noise level $\zeta=4\%$ and then use that for BM3D and KSVD. We use the pre-trained ANNs of ADNet and DnCNN given in the literature, and denoised test images with 17 layers in ADNet and 20 layers in DnCNN. Fig.~\ref{fig:compNoiL2} represents the original images, noisy images, and denoised images. Similarly to the case $\zeta=4\%$, we visually observe that EGGD and ADNet perform better than the other three methods. Table~\ref{tab} shows that the average ShE of the original image is closer to that of denoised images of EGGD and ADNet than the other three methods. Moreover, Fig.~\ref{fig:boxPlot} shows box and whisker plots for the three similarity metrics of remote sensing image denoising with EGGD and four other state-of-the-art denoising methods. The largest average PSNR and SSIM values are attained by EGGD; thus, EGGD possesses the highest numerical and structural similarity to the corresponding original image than the other three methods. 

\begin{figure*}[htp]
\centering
\includegraphics[width=1\textwidth]{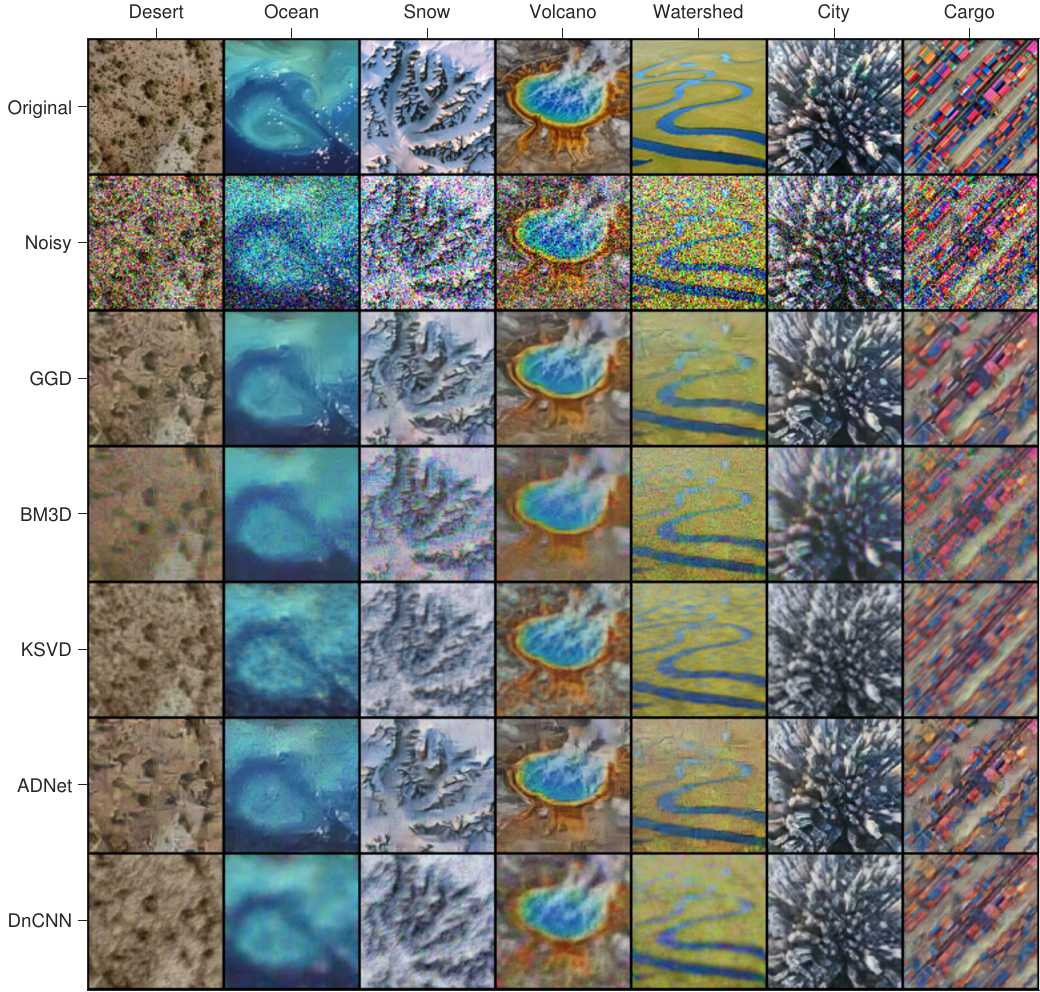}
\caption{Remote sensing image denoising with EGGD and four other state-of-the-art denoising methods, namely, sparse 3-D transform-domain collaborative filtering (BM3D), sparse and redundant representations over learned dictionaries (KSVD), Attention-guided Denoising Convolutional Neural Network (ADNet), and Denoising Convolutional Neural Networks (DnCNN). Each of the seven test images, namely, desert, ocean, snow, volcano, watershed, city, and cargo, of size $10^3\times 10^3$, is imposed with a relative Gaussian noise of 6\%. EGGD is executed with the parameter values ($\rho, \delta, L$) = (9, 20, 80), (11, 20, 80), and (11, 20, 80) for the channels Y, Cr, and Cb, respectively. The other four methods are executed with their recommended parameter values, or tested with their given trained ANNs. Here, we observe that EGGD retains texture and cartoon in the denoised images most of the time than that of the other four methods.}
\label{fig:compNoiL3}
\end{figure*}

For the noise level $\zeta=6\%$, we implement EGGD with the parameter triplets ($\rho, \delta, L$) = (9, 20, 80), (11, 20, 80), and (11, 20, 80) for the channels Y, Cr, and Cb, respectively. We arbitrarily choose these parameter values except for setting slightly bigger values like 9 and 11 for the patch size $\rho$ than that of $\zeta=4\%$, since the noise contamination in the case $\zeta=6\%$ is bigger. We implement BM3D and KSVD with the same parameter values as in the cases $\zeta=2\%$ and $\zeta=4\%$. Similar to those two previous cases, we explicitly find the corresponding $\epsilon$ for the relative noise level $\zeta=4\%$ and then use that for BM3D and KSVD. Similarly, we use the pre-trained ANNs of ADNet and DnCNN with 17 layers and 20 layers, respectively. Figure~\ref{fig:compNoiL3} represents the original images, noisy images, and denoised images of all the five methods for the noise level $\zeta=6\%$. Here, we observe that EGGD, KSVD, and ADNet perform better than the other two methods. Table~\ref{tab} shows that the average ShE of the original image is closest to that of denoised images of EGGD and ADNet, and then closer to KSVD and DnCNN. Moreover, Fig.~\ref{fig:boxPlot} shows box and whisker plots for the three similarity metrics of remote sensing image denoising with EGGD and four other state-of-the-art denoising methods. The largest average PSNR and SSIM values are attained by EGGD; thus, EGGD possesses the highest numerical and structural similarity to the corresponding original image than the other three methods.

According to the grand averages of the performance metrics, the rank of the best to the worst ShE is ADNet, EGGD, DnCNN, KSVD, and BM3D; however, the difference between the best and the worst is as small as 0.46. This order justifies the similarity of the information content of the denoised images to the corresponding original image. The ranks of the best to the worst PSNR and SSIM are EGGD, ADNet, KSVD, DnCNN, and BM3D. This rank infers the best to the worst numerical and structural similarity of the denoised images to the corresponding original images. 

\section{Discussion and Conclusion}\label{sec:con} 
The quality of remote sensing images is often degraded due to environmental factors and the issues of the imaging system, which may impair subsequent image analysis. Denoising is an essential task in remote sensing imaging; however, the current denoising algorithms fail to attain optimum performance due to the complexity of features in the texture. Here, we presented a computationally efficient and robust remote sensing image denoising framework. This technique was capable of producing accurate images without any additional training image samples like in ANN-based methods; thus, it requires only limited resources.  To preserve the smoothness and features across the image, this method non-locally worked on patches partitioned from the image. It implemented denoising based on the singular value spectrum of the geodesics' Gramian matrix of the patch space in which singular values are approximated by the randomized SVD scheme to increase efficiency. We validated the performance of this method using diverse remote sensing images contaminated with diverse noise levels. The performance is evaluated with three benchmark similarity metrics, ShE, PSNR, and SSIM.

Similar to BM3D and KSVD, EGGD is also an algebraic non-local denoising framework. BM3D stacks similar patches into 3D groups by block matching and then transforms these 3D groups into the wavelet domain, and KSVD is based on sparse and redundant representations over trained dictionaries. However, EGGD represents each patch in a high-dimensional space and then transforms this space into a low-rank manifold. BM3D utilizes hard thresholding or Wiener filtering with coefficients in the wavelet domain followed by applying an inverse transformation of coefficients, and KSVD performs denoising using a global image prior estimated by a Bayesian reconstruction framework that forces sparsity. In contrast, EGGD employs manifold learning concepts and computes prominent SVs of the Gramian matrix by randomized techniques. Each denoised version of the image of both BM3D and EGGD is constructed by aggregating all the estimated patches, which is not the practice in KSVD. Even though the increasing noise drastically decreases the denoising performance of BM3D, that of EGGD is gradual due to its manifold approach. KSVD suffers from semi-convergence as the computation of the right dictionary for a dataset is a non-convex problem; however, EGGD is free from semi-convergence as this framework doesn't include any optimization.

Apart from algebraic denoising frameworks like EGGD, ADNet and DnCNN require significantly more training data as they are ANNs. ADNet integrates local as well as global feature information and extracts the hidden noise information, and DnCNN attains denoising performance by utilizing residual learning and batch normalization process. In contrast, EGGD eliminates hidden noise via a homeomorphism projection of high-dimensional data into a low-rank manifold. DnCNN model is fabricated for the tasks with Gaussian noise contamination, whereas EGGD is tailored for any type of noise sample. As opposed to EGDD, ANN frameworks like ADNet and DnCNN may be inefficient in providing the original structural features of the remote sensing images at increasing noise levels and also require a significant amount of training samples for sustainable performance.

Color subsampling takes advantage of the fact that the human visual system has a lot of light-sensitive elements that detect brightness while it has a small number of color-sensitive elements to detect color. We observed in the experiments that ShE, PSNR, and SSIM values for the channels R, G, and B are similar while those values for the channels Chrominance-blue (Cb) and Chrominance-red (Cr) are considerably lower than that of channel Luminance (Y). Thus, denoising RGB images directly through channels R, G, and B would not be efficient since it increases computational cost as each channel is treated with the same emphasis. Moreover, minor changes in the pixel intensities on channels R, G, and B during the denoising process might change the color of the RGB image significantly. Channel Y of the YCbCr representation possesses the greatest information (measured by ShE) and is similar to the original RGB image in both pixel intensities (measured by PSNR) and structure (measured by SSIM). Thus, the transformation of the RGB image into YCbCr is preferred before denoising so that image denoising can be given different focuses for brightness and colors in which channel Y infers the brightness of the image while both Cb and Cr channels infer the color of the image. This process of denoising offers high efficiency and accuracy as more emphasis can be given to Y than that of Cb and Cr which we utilize by setting diverse values for the parameters $\rho$, $\delta$, and $L$ across channels. Moreover, since remote sensing images contain more information than regular images, better emphasis can also be given to both Cb and Cr channels for the superior quality of image denoising. 

In contrast to the state-of-the-art denoising methods, EGGD possesses only three parameters, patch size, nearest neighborhood size, and SV threshold, that can easily be pre-determined. As we have seen in Sec.~\ref{sec:exa}, the nearest neighborhood size and SV threshold have less influence on denoising performance. Thus, we can set them into fixed general values as we have done in Sec.~\ref{sec:exa}. The patch size should be set such that a bigger value for bigger noise and vice versa. We set the next value of the patch sizes for Cb and Cr bigger than that of Y since the noise contamination in Cb and Cr is less than that of Y. In the future, we will utilize a trial-and-error process to derive the ideal patch size for the channels Cb and Cr based on that of Y. Moreover, we will also perform an analysis to predefine the parameter values for the nearest neighborhood size and the SV threshold.

Hyperparameter tuning is a rigorous task and a limitation in the model fitting of data-driven processes including our EGGD framework. Thus, apart from the procedures explained above to determine the best hyperparameter values, we will also be planning to implement a Bayesian parameter estimation routine to determine the best hyperparameter values. Bayesian parameter estimation is a widely used technique for estimating the probability density function of random variables with unknown parameters. In our case, while the random variable is a similarity measure such as PSNR, the parameters are patch size ($\rho$), nearest neighborhood size ($\delta$), and SV threshold ($L$). With some possible parameter space of $\{(\rho, \delta, L) \ \vert \ \rho = 1, 3, 5,\dots, 15; \delta = 2, 4,\dots, 40; L=10, 20, \dots, 200\}$, the probability density function of PSNR for a given value of $(\rho, \delta, L)$ is denoted by $p(PSNR\vert(\rho, \delta, L))$. Then, we will draw a set of independent samples $S = \{(\rho_1, \delta_1, L_1), \dots, (\rho_n, \delta_n, L_n)\}$ from a remote sensing denoising experiment. Our goal is to compute $p(PSNR\vert S)$ as close as knowing the unknown probability density function of PSNR, i.e., $p(\rho, \delta, L)$. We will determine the parameter combination $(\rho, \delta, L)$ for the best PSNR and will utilize that for denoising the remote sensing images to assure accuracy and robustness.

The complete EGGD framework consists of three components, SV approximation, single-channel denoising, and three-channel extension, in which SV approximation consumes most of the complexity. For a given RGB image of size $n\times n\times 3$, an intermediate step of EGGD requires the computation of prominent SVs of a Gramian matrix of size $n^2\times n^2$. While the regular SVD to compute SVs of the denoising process of an image of size $n\times n\times 3$ consumes a computational complexity of $\mathcal{O}(n^6)$ \citep{gajamannage2022efficient}, the SV approximation framework RSVD consumes only a complexity of $\mathcal{O}\left(n^4 + k^2n^2\right)$ where $k$ is the number of randomized SVs of interest. Since EGGD performs accurate denoising with a significantly small number of prominent SVs, $k$ is significantly less than $n$. Thus, the ultimate computational complexity of EGGD becomes $\mathcal{O}(n^4)$ as $k^2n^2$ is less influential for the final complexity. Thus, this randomized approach for SVD reduces the computational complexity of EGGD by $ n^2$ times. Especially, in our choice of remote sensing images of size $n=10^3$, the randomized approach reduces the complexity by $10^6$ times than that of using the regular SVD.

The implementation of the randomized SVD instead of the regular SVD makes EGGD significantly efficient. However, the computational complexity of EGGD is still $\mathcal{O}(n^4)$ for a square-shaped image of length $n$-pixels. This computational cost can further be reduced by an artificial neural network surrogate, instead of the current linear algebraic approach, which we propose as future work. We plan to train this neural network on a wide corpus of remote sensing images available in open-source databases. The training of this neural network is rigorous due to the scale of the data; however, the training is only a one-time process for a required level of accuracy. This trained model can be used to denoise noisy remote sensing images more efficiently.

Here, we proposed an efficient and robust remote sensing image denoising framework which we tested on seven benchmark test images imposed with three Gaussian noise levels and compared with four (two algebraic and two ANN-based) state-of-the-art methods. The results justify the superiority of the proposed method in remote sensing image denoising. 

\section*{Data Availability Statement}
Data will be made available on request from the corresponding author.

\section*{Disclosure statement}
No potential conflict of interest was reported by the authors.

\section*{Funding}
This research was partially supported by the National Science Foundation under grant number DMS-2418826.


\begin{thebibliography}{}

\bibitem [\protect \citeauthoryear {%
Agarwal%
\ \BBA {} Erickson%
}{%
Agarwal%
\ \BBA {} Erickson%
}{%
{\protect \APACyear {1997}}%
}]{%
agarwal1999geometric}
\APACinsertmetastar {%
agarwal1999geometric}%
\begin{APACrefauthors}%
Agarwal, P\BPBI K.%
\BCBT {}\ \BBA {} Erickson, J.%
\end{APACrefauthors}%
\unskip\
\newblock
\APACrefYearMonthDay{1997}{}{}.
\newblock
{\BBOQ}\APACrefatitle {{Geometric range searching and its relatives}}
  {{Geometric range searching and its relatives}}.{\BBCQ}
\newblock
\APACjournalVolNumPages{Advances in Discrete and Computational
  Geometry}{223}{}{1--56}.
\newblock
\begin{APACrefDOI} \doi{10.1.1.38.6261} \end{APACrefDOI}
\PrintBackRefs{\CurrentBib}

\bibitem [\protect \citeauthoryear {%
Campbell%
\ \BBA {} Wynne%
}{%
Campbell%
\ \BBA {} Wynne%
}{%
{\protect \APACyear {2011}}%
}]{%
campbell2011introduction}
\APACinsertmetastar {%
campbell2011introduction}%
\begin{APACrefauthors}%
Campbell, J\BPBI B.%
\BCBT {}\ \BBA {} Wynne, R\BPBI H.%
\end{APACrefauthors}%
\unskip\
\newblock
\APACrefYear{2011}.
\newblock
\APACrefbtitle {Introduction to remote sensing} {Introduction to remote
  sensing}.
\newblock
\APACaddressPublisher{}{Guilford press}.
\PrintBackRefs{\CurrentBib}

\bibitem [\protect \citeauthoryear {%
Cover%
}{%
Cover%
}{%
{\protect \APACyear {1999}}%
}]{%
cover1999elements}
\APACinsertmetastar {%
cover1999elements}%
\begin{APACrefauthors}%
Cover, T\BPBI M.%
\end{APACrefauthors}%
\unskip\
\newblock
\APACrefYear{1999}.
\newblock
\APACrefbtitle {Elements of information theory} {Elements of information
  theory}.
\newblock
\APACaddressPublisher{}{John Wiley \& Sons}.
\PrintBackRefs{\CurrentBib}

\bibitem [\protect \citeauthoryear {%
Dabov%
, Foi%
, Katkovnik%
\BCBL {}\ \BBA {} Egiazarian%
}{%
Dabov%
\ \protect \BOthers {.}}{%
{\protect \APACyear {2007}}%
}]{%
Dabov2007}
\APACinsertmetastar {%
Dabov2007}%
\begin{APACrefauthors}%
Dabov, K.%
, Foi, A.%
, Katkovnik, V.%
\BCBL {}\ \BBA {} Egiazarian, K.%
\end{APACrefauthors}%
\unskip\
\newblock
\APACrefYearMonthDay{2007}{aug}{}.
\newblock
{\BBOQ}\APACrefatitle {{Image denoising by sparse 3-D transform-domain
  collaborative filtering}} {{Image denoising by sparse 3-D transform-domain
  collaborative filtering}}.{\BBCQ}
\newblock
\APACjournalVolNumPages{IEEE Transactions on Image
  Processing}{16}{8}{2080--2095}.
\newblock
\begin{APACrefDOI} \doi{10.1109/TIP.2007.901238} \end{APACrefDOI}
\PrintBackRefs{\CurrentBib}

\bibitem [\protect \citeauthoryear {%
Elad%
\ \BBA {} Aharon%
}{%
Elad%
\ \BBA {} Aharon%
}{%
{\protect \APACyear {2006}}%
}]{%
Elad2006}
\APACinsertmetastar {%
Elad2006}%
\begin{APACrefauthors}%
Elad, M.%
\BCBT {}\ \BBA {} Aharon, M.%
\end{APACrefauthors}%
\unskip\
\newblock
\APACrefYearMonthDay{2006}{dec}{}.
\newblock
{\BBOQ}\APACrefatitle {{Image denoising via sparse and redundant
  representations over learned dictionaries}} {{Image denoising via sparse and
  redundant representations over learned dictionaries}}.{\BBCQ}
\newblock
\APACjournalVolNumPages{IEEE Transactions on Image
  Processing}{15}{12}{3736--3745}.
\newblock
\begin{APACrefDOI} \doi{10.1109/TIP.2006.881969} \end{APACrefDOI}
\PrintBackRefs{\CurrentBib}

\bibitem [\protect \citeauthoryear {%
Fan%
, Zhang%
, Fan%
\BCBL {}\ \BBA {} Zhang%
}{%
Fan%
\ \protect \BOthers {.}}{%
{\protect \APACyear {2019}}%
}]{%
Fan2019}
\APACinsertmetastar {%
Fan2019}%
\begin{APACrefauthors}%
Fan, L.%
, Zhang, F.%
, Fan, H.%
\BCBL {}\ \BBA {} Zhang, C.%
\end{APACrefauthors}%
\unskip\
\newblock
\APACrefYearMonthDay{2019}{}{}.
\newblock
{\BBOQ}\APACrefatitle {{Brief review of image denoising techniques}} {{Brief
  review of image denoising techniques}}.{\BBCQ}
\newblock
\APACjournalVolNumPages{Visual Computing for Industry, Biomedicine, and
  Art}{2}{1}{1--12}.
\newblock
\begin{APACrefDOI} \doi{10.1186/s42492-019-0016-7} \end{APACrefDOI}
\PrintBackRefs{\CurrentBib}

\bibitem [\protect \citeauthoryear {%
Floyd%
}{%
Floyd%
}{%
{\protect \APACyear {1962}}%
}]{%
floyd1962algorithm}
\APACinsertmetastar {%
floyd1962algorithm}%
\begin{APACrefauthors}%
Floyd, R\BPBI W.%
\end{APACrefauthors}%
\unskip\
\newblock
\APACrefYearMonthDay{1962}{}{}.
\newblock
{\BBOQ}\APACrefatitle {{Algorithm 97: shortest path}} {{Algorithm 97: shortest
  path}}.{\BBCQ}
\newblock
\APACjournalVolNumPages{Communications of the ACM}{5}{6}{345}.
\newblock
\begin{APACrefDOI} \doi{10.1145/367766.368168} \end{APACrefDOI}
\PrintBackRefs{\CurrentBib}

\bibitem [\protect \citeauthoryear {%
Gajamannage%
, Butail%
, Porfiri%
\BCBL {}\ \BBA {} Bollt%
}{%
Gajamannage%
\ \protect \BOthers {.}}{%
{\protect \APACyear {2015}}%
{\protect \APACexlab {{\protect \BCnt {1}}}}}]{%
gajamannage2015a}
\APACinsertmetastar {%
gajamannage2015a}%
\begin{APACrefauthors}%
Gajamannage, K.%
, Butail, S.%
, Porfiri, M.%
\BCBL {}\ \BBA {} Bollt, E\BPBI M.%
\end{APACrefauthors}%
\unskip\
\newblock
\APACrefYearMonthDay{2015{\protect \BCnt {1}}}{jan}{}.
\newblock
{\BBOQ}\APACrefatitle {{Dimensionality reduction of collective motion by
  principal manifolds}} {{Dimensionality reduction of collective motion by
  principal manifolds}}.{\BBCQ}
\newblock
\APACjournalVolNumPages{Physica D: Nonlinear Phenomena}{291}{}{62--73}.
\newblock
\begin{APACrefDOI} \doi{10.1016/j.physd.2014.09.009} \end{APACrefDOI}
\PrintBackRefs{\CurrentBib}

\bibitem [\protect \citeauthoryear {%
Gajamannage%
, Butail%
, Porfiri%
\BCBL {}\ \BBA {} Bollt%
}{%
Gajamannage%
\ \protect \BOthers {.}}{%
{\protect \APACyear {2015}}%
{\protect \APACexlab {{\protect \BCnt {2}}}}}]{%
gajamannage2015identifying}
\APACinsertmetastar {%
gajamannage2015identifying}%
\begin{APACrefauthors}%
Gajamannage, K.%
, Butail, S.%
, Porfiri, M.%
\BCBL {}\ \BBA {} Bollt, E\BPBI M.%
\end{APACrefauthors}%
\unskip\
\newblock
\APACrefYearMonthDay{2015{\protect \BCnt {2}}}{}{}.
\newblock
{\BBOQ}\APACrefatitle {Identifying manifolds underlying group motion in Vicsek
  agents} {Identifying manifolds underlying group motion in vicsek
  agents}.{\BBCQ}
\newblock
\APACjournalVolNumPages{The European Physical Journal Special
  Topics}{224}{}{3245--3256}.
\PrintBackRefs{\CurrentBib}

\bibitem [\protect \citeauthoryear {%
Gajamannage%
\ \BBA {} Paffenroth%
}{%
Gajamannage%
\ \BBA {} Paffenroth%
}{%
{\protect \APACyear {2021}}%
}]{%
gajamannage2021}
\APACinsertmetastar {%
gajamannage2021}%
\begin{APACrefauthors}%
Gajamannage, K.%
\BCBT {}\ \BBA {} Paffenroth, R.%
\end{APACrefauthors}%
\unskip\
\newblock
\APACrefYearMonthDay{2021}{dec}{}.
\newblock
{\BBOQ}\APACrefatitle {{Bounded manifold completion}} {{Bounded manifold
  completion}}.{\BBCQ}
\newblock
\APACjournalVolNumPages{Pattern Recognition}{111}{}{107661}.
\newblock
\begin{APACrefDOI} \doi{https://doi.org/10.1016/j.patcog.2020.107661}
  \end{APACrefDOI}
\PrintBackRefs{\CurrentBib}

\bibitem [\protect \citeauthoryear {%
Gajamannage%
, Paffenroth%
\BCBL {}\ \BBA {} Bollt%
}{%
Gajamannage%
\ \protect \BOthers {.}}{%
{\protect \APACyear {2019}}%
}]{%
SGE}
\APACinsertmetastar {%
SGE}%
\begin{APACrefauthors}%
Gajamannage, K.%
, Paffenroth, R.%
\BCBL {}\ \BBA {} Bollt, E\BPBI M.%
\end{APACrefauthors}%
\unskip\
\newblock
\APACrefYearMonthDay{2019}{mar}{}.
\newblock
{\BBOQ}\APACrefatitle {{A nonlinear dimensionality reduction framework using
  smooth geodesics}} {{A nonlinear dimensionality reduction framework using
  smooth geodesics}}.{\BBCQ}
\newblock
\APACjournalVolNumPages{Pattern Recognition}{87}{}{226--236}.
\newblock
\begin{APACrefDOI} \doi{10.1016/j.patcog.2018.10.020} \end{APACrefDOI}
\PrintBackRefs{\CurrentBib}

\bibitem [\protect \citeauthoryear {%
Gajamannage%
, Paffenroth%
\BCBL {}\ \BBA {} Jayasumana%
}{%
Gajamannage%
\ \protect \BOthers {.}}{%
{\protect \APACyear {2024}}%
}]{%
gajamannage2020patch}
\APACinsertmetastar {%
gajamannage2020patch}%
\begin{APACrefauthors}%
Gajamannage, K.%
, Paffenroth, R.%
\BCBL {}\ \BBA {} Jayasumana, A\BPBI P.%
\end{APACrefauthors}%
\unskip\
\newblock
\APACrefYearMonthDay{2024}{}{}.
\newblock
{\BBOQ}\APACrefatitle {{Image Denoising Using the Geodesics' Gramian of the Manifold Underlying Patch-Space}} {{Image Denoising Using the Geodesics' Gramian of the Manifold Underlying Patch-Space}}.{\BBCQ}
\newblock
\APACjournalVolNumPages{Statistics, Optimization \& Information Computing}{12}{6}{1775--1794}.
\newblock
\begin{APACrefDOI} \doi{10.19139/soic-2310-5070-2124} \end{APACrefDOI}
\PrintBackRefs{\CurrentBib}

\bibitem [\protect \citeauthoryear {%
Gajamannage%
, Park%
, Muddamallappa%
\BCBL {}\ \BBA {} Mathur%
}{%
Gajamannage%
, Park%
, Muddamallappa%
\BCBL {}\ \BBA {} Mathur%
}{%
{\protect \APACyear {2022}}%
}]{%
gajamannage2022efficient}
\APACinsertmetastar {%
gajamannage2022efficient}%
\begin{APACrefauthors}%
Gajamannage, K.%
, Park, Y.%
, Muddamallappa, M.%
\BCBL {}\ \BBA {} Mathur, S.%
\end{APACrefauthors}%
\unskip\
\newblock
\APACrefYearMonthDay{2022}{}{}.
\newblock
{\BBOQ}\APACrefatitle {Efficient noise filtration of images by low-rank
  singular vector approximations of Geodesics' Gramian Matrix} {Efficient noise
  filtration of images by low-rank singular vector approximations of geodesics'
  gramian matrix}.{\BBCQ}
\newblock
\APACjournalVolNumPages{arXiv preprint arXiv:2209.13094}{}{}{}.
\PrintBackRefs{\CurrentBib}

\bibitem [\protect \citeauthoryear {%
Gajamannage%
, Park%
, Paffenroth%
\BCBL {}\ \BBA {} Jayasumana%
}{%
Gajamannage%
, Park%
, Paffenroth%
\BCBL {}\ \BBA {} Jayasumana%
}{%
{\protect \APACyear {2022}}%
}]{%
gajamannage2022reconstruction}
\APACinsertmetastar {%
gajamannage2022reconstruction}%
\begin{APACrefauthors}%
Gajamannage, K.%
, Park, Y.%
, Paffenroth, R.%
\BCBL {}\ \BBA {} Jayasumana, A\BPBI P.%
\end{APACrefauthors}%
\unskip\
\newblock
\APACrefYearMonthDay{2022}{nov}{}.
\newblock
{\BBOQ}\APACrefatitle {{Reconstruction of fragmented trajectories of collective
  motion using Hadamard deep autoencoders}} {{Reconstruction of fragmented
  trajectories of collective motion using Hadamard deep autoencoders}}.{\BBCQ}
\newblock
\APACjournalVolNumPages{Pattern Recognition}{131}{}{108891}.
\newblock
\begin{APACrefDOI} \doi{10.1016/j.patcog.2022.108891} \end{APACrefDOI}
\PrintBackRefs{\CurrentBib}

\bibitem [\protect \citeauthoryear {%
Gajamannage%
, Park%
\BCBL {}\ \BBA {} Sadovski%
}{%
Gajamannage%
, Park%
\BCBL {}\ \BBA {} Sadovski%
}{%
{\protect \APACyear {2022}}%
}]{%
GajamannageGGDwtNoise}
\APACinsertmetastar {%
GajamannageGGDwtNoise}%
\begin{APACrefauthors}%
Gajamannage, K.%
, Park, Y.%
\BCBL {}\ \BBA {} Sadovski, A.%
\end{APACrefauthors}%
\unskip\
\newblock
\APACrefYearMonthDay{2022}{}{}.
\newblock
{\BBOQ}\APACrefatitle {{Geodesic Gramian Denoising Applied to the Images
  Contaminated With Noise Sampled From Diverse Probability Distributions}}
  {{Geodesic Gramian Denoising Applied to the Images Contaminated With Noise
  Sampled From Diverse Probability Distributions}}.{\BBCQ}
\newblock
\APACjournalVolNumPages{IET Image Processing}{00}{}{1--13}.
\newblock
\begin{APACrefDOI} \doi{10.1049/ipr2.12623} \end{APACrefDOI}
\PrintBackRefs{\CurrentBib}

\bibitem [\protect \citeauthoryear {%
Golub%
\ \BBA {} Reinsch%
}{%
Golub%
\ \BBA {} Reinsch%
}{%
{\protect \APACyear {1970}}%
}]{%
Golub1970}
\APACinsertmetastar {%
Golub1970}%
\begin{APACrefauthors}%
Golub, G\BPBI H.%
\BCBT {}\ \BBA {} Reinsch, C.%
\end{APACrefauthors}%
\unskip\
\newblock
\APACrefYearMonthDay{1970}{}{}.
\newblock
\APACrefbtitle {{Handbook Series Linear Algebra Singular Value Decomposition
  and Least Squares Solutions*}} {{Handbook Series Linear Algebra Singular
  Value Decomposition and Least Squares Solutions*}}\
  \APACbVolEdTR{\BVOL~14}{\BTR{}}.
\PrintBackRefs{\CurrentBib}

\bibitem [\protect \citeauthoryear {%
Halko%
, Martinsson%
\BCBL {}\ \BBA {} Tropp%
}{%
Halko%
\ \protect \BOthers {.}}{%
{\protect \APACyear {2011}}%
}]{%
halko2011finding}
\APACinsertmetastar {%
halko2011finding}%
\begin{APACrefauthors}%
Halko, N.%
, Martinsson, P\BPBI G.%
\BCBL {}\ \BBA {} Tropp, J\BPBI A.%
\end{APACrefauthors}%
\unskip\
\newblock
\APACrefYearMonthDay{2011}{may}{}.
\newblock
{\BBOQ}\APACrefatitle {{Finding structure with randomness: Probabilistic
  algorithms for constructing approximate matrix decompositions}} {{Finding
  structure with randomness: Probabilistic algorithms for constructing
  approximate matrix decompositions}}.{\BBCQ}
\newblock
\APACjournalVolNumPages{SIAM Review}{53}{2}{217--288}.
\newblock
\begin{APACrefURL} \url{https://epubs.siam.org/doi/10.1137/090771806}
  \end{APACrefURL}
\newblock
\begin{APACrefDOI} \doi{10.1137/090771806} \end{APACrefDOI}
\PrintBackRefs{\CurrentBib}

\bibitem [\protect \citeauthoryear {%
Hore%
\ \BBA {} Ziou%
}{%
Hore%
\ \BBA {} Ziou%
}{%
{\protect \APACyear {2010}}%
}]{%
hore2010image}
\APACinsertmetastar {%
hore2010image}%
\begin{APACrefauthors}%
Hore, A.%
\BCBT {}\ \BBA {} Ziou, D.%
\end{APACrefauthors}%
\unskip\
\newblock
\APACrefYearMonthDay{2010}{}{}.
\newblock
{\BBOQ}\APACrefatitle {Image quality metrics: PSNR vs. SSIM} {Image quality
  metrics: Psnr vs. ssim}.{\BBCQ}
\newblock
\BIn{} \APACrefbtitle {2010 20th international conference on pattern
  recognition} {2010 20th international conference on pattern recognition}\
  (\BPGS\ 2366--2369).
\PrintBackRefs{\CurrentBib}

\bibitem [\protect \citeauthoryear {%
Kirik%
, Iskandarov%
, Erturk%
\BCBL {}\ \BBA {} Ozdemir%
}{%
Kirik%
\ \protect \BOthers {.}}{%
{\protect \APACyear {2024}}%
}]{%
kirik2024quantitative}
\APACinsertmetastar {%
kirik2024quantitative}%
\begin{APACrefauthors}%
Kirik, F.%
, Iskandarov, F.%
, Erturk, K\BPBI M.%
\BCBL {}\ \BBA {} Ozdemir, H.%
\end{APACrefauthors}%
\unskip\
\newblock
\APACrefYearMonthDay{2024}{}{}.
\newblock
{\BBOQ}\APACrefatitle {Quantitative analysis of deep learning-based denoising
  model efficacy on optical coherence tomography images with different noise
  levels} {Quantitative analysis of deep learning-based denoising model
  efficacy on optical coherence tomography images with different noise
  levels}.{\BBCQ}
\newblock
\APACjournalVolNumPages{Photodiagnosis and Photodynamic Therapy}{45}{}{103891}.
\PrintBackRefs{\CurrentBib}

\bibitem [\protect \citeauthoryear {%
Kusangaya%
\ \BBA {} Sithole%
}{%
Kusangaya%
\ \BBA {} Sithole%
}{%
{\protect \APACyear {2015}}%
}]{%
kusangaya2015remote}
\APACinsertmetastar {%
kusangaya2015remote}%
\begin{APACrefauthors}%
Kusangaya, S.%
\BCBT {}\ \BBA {} Sithole, V\BPBI B.%
\end{APACrefauthors}%
\unskip\
\newblock
\APACrefYearMonthDay{2015}{}{}.
\newblock
{\BBOQ}\APACrefatitle {Remote sensing-based fire frequency mapping in a
  savannah rangeland} {Remote sensing-based fire frequency mapping in a
  savannah rangeland}.{\BBCQ}
\newblock
\APACjournalVolNumPages{South African Journal of Geomatics}{4}{1}{36--49}.
\PrintBackRefs{\CurrentBib}

\bibitem [\protect \citeauthoryear {%
Lee%
, Lendasse%
\BCBL {}\ \BBA {} Verleysen%
}{%
Lee%
\ \protect \BOthers {.}}{%
{\protect \APACyear {2004}}%
}]{%
lee2004nonlinear}
\APACinsertmetastar {%
lee2004nonlinear}%
\begin{APACrefauthors}%
Lee, J\BPBI A.%
, Lendasse, A.%
\BCBL {}\ \BBA {} Verleysen, M.%
\end{APACrefauthors}%
\unskip\
\newblock
\APACrefYearMonthDay{2004}{}{}.
\newblock
{\BBOQ}\APACrefatitle {{Nonlinear projection with curvilinear distances: Isomap
  versus curvilinear distance analysis}} {{Nonlinear projection with
  curvilinear distances: Isomap versus curvilinear distance analysis}}.{\BBCQ}
\newblock
\APACjournalVolNumPages{Neurocomputing}{57}{1-4}{49--76}.
\newblock
\begin{APACrefDOI} \doi{10.1016/j.neucom.2004.01.007} \end{APACrefDOI}
\PrintBackRefs{\CurrentBib}

\bibitem [\protect \citeauthoryear {%
Liu%
, Wang%
, Wang%
\BCBL {}\ \BBA {} Han%
}{%
Liu%
\ \protect \BOthers {.}}{%
{\protect \APACyear {2019}}%
}]{%
liu2019remote}
\APACinsertmetastar {%
liu2019remote}%
\begin{APACrefauthors}%
Liu, P.%
, Wang, M.%
, Wang, L.%
\BCBL {}\ \BBA {} Han, W.%
\end{APACrefauthors}%
\unskip\
\newblock
\APACrefYearMonthDay{2019}{}{}.
\newblock
{\BBOQ}\APACrefatitle {Remote-sensing image denoising with multi-sourced
  information} {Remote-sensing image denoising with multi-sourced
  information}.{\BBCQ}
\newblock
\APACjournalVolNumPages{IEEE Journal of Selected Topics in Applied Earth
  Observations and Remote Sensing}{12}{2}{660--674}.
\PrintBackRefs{\CurrentBib}

\bibitem [\protect \citeauthoryear {%
Lossou%
, Owusu-Prempeh%
\BCBL {}\ \BBA {} Agyemang%
}{%
Lossou%
\ \protect \BOthers {.}}{%
{\protect \APACyear {2019}}%
}]{%
lossou2019monitoring}
\APACinsertmetastar {%
lossou2019monitoring}%
\begin{APACrefauthors}%
Lossou, E.%
, Owusu-Prempeh, N.%
\BCBL {}\ \BBA {} Agyemang, G.%
\end{APACrefauthors}%
\unskip\
\newblock
\APACrefYearMonthDay{2019}{}{}.
\newblock
{\BBOQ}\APACrefatitle {Monitoring Land Cover changes in the tropical high
  forests using multi-temporal remote sensing and spatial analysis techniques}
  {Monitoring land cover changes in the tropical high forests using
  multi-temporal remote sensing and spatial analysis techniques}.{\BBCQ}
\newblock
\APACjournalVolNumPages{Remote Sensing Applications: Society and
  Environment}{16}{}{100264}.
\PrintBackRefs{\CurrentBib}

\bibitem [\protect \citeauthoryear {%
Matsunobu%
, Pedro%
\BCBL {}\ \BBA {} Coimbra%
}{%
Matsunobu%
\ \protect \BOthers {.}}{%
{\protect \APACyear {2021}}%
}]{%
matsunobu2021cloud}
\APACinsertmetastar {%
matsunobu2021cloud}%
\begin{APACrefauthors}%
Matsunobu, L\BPBI M.%
, Pedro, H\BPBI T.%
\BCBL {}\ \BBA {} Coimbra, C\BPBI F.%
\end{APACrefauthors}%
\unskip\
\newblock
\APACrefYearMonthDay{2021}{}{}.
\newblock
{\BBOQ}\APACrefatitle {Cloud detection using convolutional neural networks on
  remote sensing images} {Cloud detection using convolutional neural networks
  on remote sensing images}.{\BBCQ}
\newblock
\APACjournalVolNumPages{Solar Energy}{230}{}{1020--1032}.
\PrintBackRefs{\CurrentBib}

\bibitem [\protect \citeauthoryear {%
Meyer%
\ \BBA {} Shen%
}{%
Meyer%
\ \BBA {} Shen%
}{%
{\protect \APACyear {2014}}%
}]{%
Meyer2014}
\APACinsertmetastar {%
Meyer2014}%
\begin{APACrefauthors}%
Meyer, F\BPBI G.%
\BCBT {}\ \BBA {} Shen, X.%
\end{APACrefauthors}%
\unskip\
\newblock
\APACrefYearMonthDay{2014}{mar}{}.
\newblock
\APACrefbtitle {{Perturbation of the eigenvectors of the graph Laplacian:
  Application to image denoising}} {{Perturbation of the eigenvectors of the
  graph Laplacian: Application to image denoising}}\ (\BVOL~36)\ (\BNUM~2).
\newblock
\APACaddressPublisher{}{Academic Press Inc.}
\newblock
\begin{APACrefDOI} \doi{10.1016/j.acha.2013.06.004} \end{APACrefDOI}
\PrintBackRefs{\CurrentBib}

\bibitem [\protect \citeauthoryear {%
Rubinstein%
, Bruckstein%
\BCBL {}\ \BBA {} Elad%
}{%
Rubinstein%
\ \protect \BOthers {.}}{%
{\protect \APACyear {2010}}%
}]{%
Rubinstein2010}
\APACinsertmetastar {%
Rubinstein2010}%
\begin{APACrefauthors}%
Rubinstein, R.%
, Bruckstein, A\BPBI M.%
\BCBL {}\ \BBA {} Elad, M.%
\end{APACrefauthors}%
\unskip\
\newblock
\APACrefYearMonthDay{2010}{}{}.
\newblock
{\BBOQ}\APACrefatitle {{Dictionaries for sparse representation modeling}}
  {{Dictionaries for sparse representation modeling}}.{\BBCQ}
\newblock
\APACjournalVolNumPages{Proceedings of the IEEE}{98}{6}{1045--1057}.
\newblock
\begin{APACrefDOI} \doi{10.1109/JPROC.2010.2040551} \end{APACrefDOI}
\PrintBackRefs{\CurrentBib}

\bibitem [\protect \citeauthoryear {%
Shannon%
\ \protect \BOthers {.}}{%
Shannon%
\ \protect \BOthers {.}}{%
{\protect \APACyear {1959}}%
}]{%
shannon1959coding}
\APACinsertmetastar {%
shannon1959coding}%
\begin{APACrefauthors}%
Shannon, C\BPBI E.%
\BCBT {}\ \BOthersPeriod {.}
\end{APACrefauthors}%
\unskip\
\newblock
\APACrefYearMonthDay{1959}{}{}.
\newblock
{\BBOQ}\APACrefatitle {Coding theorems for a discrete source with a fidelity
  criterion} {Coding theorems for a discrete source with a fidelity
  criterion}.{\BBCQ}
\newblock
\APACjournalVolNumPages{IRE Nat. Conv. Rec}{4}{142-163}{1}.
\PrintBackRefs{\CurrentBib}

\bibitem [\protect \citeauthoryear {%
Shepard%
}{%
Shepard%
}{%
{\protect \APACyear {1968}}%
}]{%
Shepard1968}
\APACinsertmetastar {%
Shepard1968}%
\begin{APACrefauthors}%
Shepard, D.%
\end{APACrefauthors}%
\unskip\
\newblock
\APACrefYearMonthDay{1968}{jan}{}.
\newblock
{\BBOQ}\APACrefatitle {{A two-dimensional interpolation function for
  irregularly-spaced data}} {{A two-dimensional interpolation function for
  irregularly-spaced data}}.{\BBCQ}
\newblock
\BIn{} \APACrefbtitle {Proceedings of the 1968 23rd ACM National Conference,
  ACM 1968} {Proceedings of the 1968 23rd acm national conference, acm 1968}\
  (\BPGS\ 517--524).
\newblock
\APACaddressPublisher{New York, New York, USA}{ACM Press}.
\newblock
\begin{APACrefDOI} \doi{10.1145/800186.810616} \end{APACrefDOI}
\PrintBackRefs{\CurrentBib}

\bibitem [\protect \citeauthoryear {%
Tian%
\ \protect \BOthers {.}}{%
Tian%
\ \protect \BOthers {.}}{%
{\protect \APACyear {2020}}%
}]{%
tian2020attention}
\APACinsertmetastar {%
tian2020attention}%
\begin{APACrefauthors}%
Tian, C.%
, Xu, Y.%
, Li, Z.%
, Zuo, W.%
, Fei, L.%
\BCBL {}\ \BBA {} Liu, H.%
\end{APACrefauthors}%
\unskip\
\newblock
\APACrefYearMonthDay{2020}{}{}.
\newblock
{\BBOQ}\APACrefatitle {Attention-guided CNN for image denoising}
  {Attention-guided cnn for image denoising}.{\BBCQ}
\newblock
\APACjournalVolNumPages{Neural Networks}{124}{}{117--129}.
\PrintBackRefs{\CurrentBib}

\bibitem [\protect \citeauthoryear {%
Wang%
, Bovik%
, Sheikh%
\BCBL {}\ \BBA {} Simoncelli%
}{%
Wang%
\ \protect \BOthers {.}}{%
{\protect \APACyear {2004}}%
}]{%
wang2004image}
\APACinsertmetastar {%
wang2004image}%
\begin{APACrefauthors}%
Wang, Z.%
, Bovik, A\BPBI C.%
, Sheikh, H\BPBI R.%
\BCBL {}\ \BBA {} Simoncelli, E\BPBI P.%
\end{APACrefauthors}%
\unskip\
\newblock
\APACrefYearMonthDay{2004}{}{}.
\newblock
{\BBOQ}\APACrefatitle {Image quality assessment: from error visibility to
  structural similarity} {Image quality assessment: from error visibility to
  structural similarity}.{\BBCQ}
\newblock
\APACjournalVolNumPages{IEEE transactions on image
  processing}{13}{4}{600--612}.
\PrintBackRefs{\CurrentBib}

\bibitem [\protect \citeauthoryear {%
Zhang%
, Zuo%
, Chen%
, Meng%
\BCBL {}\ \BBA {} Zhang%
}{%
Zhang%
\ \protect \BOthers {.}}{%
{\protect \APACyear {2017}}%
}]{%
zhang2017beyond}
\APACinsertmetastar {%
zhang2017beyond}%
\begin{APACrefauthors}%
Zhang, K.%
, Zuo, W.%
, Chen, Y.%
, Meng, D.%
\BCBL {}\ \BBA {} Zhang, L.%
\end{APACrefauthors}%
\unskip\
\newblock
\APACrefYearMonthDay{2017}{}{}.
\newblock
{\BBOQ}\APACrefatitle {Beyond a gaussian denoiser: Residual learning of deep
  cnn for image denoising} {Beyond a gaussian denoiser: Residual learning of
  deep cnn for image denoising}.{\BBCQ}
\newblock
\APACjournalVolNumPages{IEEE transactions on image
  processing}{26}{7}{3142--3155}.
\PrintBackRefs{\CurrentBib}

\end{thebibliography}

\end{document}